\UseRawInputEncoding

\documentclass[10pt, prb, aps, twocolumn, showpacs, citeautoscript, floatfix, reprint, amsmath, amssymb, notitlepage, superscriptaddress]{revtex4-1}

\usepackage{graphicx}
\usepackage{stmaryrd}
\usepackage{rotating}
\usepackage{amsmath}
\usepackage{amsfonts}
\usepackage{amssymb}
\usepackage[countmax]{subfloat}
\usepackage{dcolumn} % align table columns on decimal point
\usepackage{bm} % bold math
\usepackage{color}

\usepackage{float}
\usepackage{latexsym}

\begin{document}

\title{Universal topological marker}

%\title{Universal local and nonlocal topological markers}

%\title{Universal topological marker for any dimension and symmetry class}

%\title{Mapping topological order and topological quantum criticality on lattice sites by a universal topological marker}

\author{Wei Chen}

\affiliation{Department of Physics, PUC-Rio, 22451-900 Rio de Janeiro, Brazil}

\date{\rm\today}

\begin{abstract}

We elaborate that for topological insulators and topological superconductors described by Dirac models in any dimension and symmetry class, the topological order can be mapped to lattice sites by a universal topological marker. Deriving from a recently discovered momentum-space universal topological invariant, we introduce a topological operator that consists of alternating projectors to filled and empty lattice eigenstates and the position operators, multiplied by the Dirac matrices that are omitted in the Hamiltonian. The topological operator projected to lattice sites yields the topological marker, whose form is explicitly constructed for every topologically nontrivial symmetry class from 1D to 3D. The off-diagonal elements of the topological operator yields a nonlocal topological marker, which decays with a correlation length that diverges at topological phase transitions, and represents a Wannier state correlation function. Various prototype examples, including Su-Schrieffer-Heeger model, Majorana chain, Chern insulators, Bernevig-Hughes-Zhang model, 2D chiral and helical $p$-wave superconductors, lattice model of $^{3}$He B-phase, and 3D time-reversal symmetric topological insulators, etc, are employed to demonstrate the ubiquity of our formalism.

\end{abstract}

\maketitle

\section{Introduction}

The celebrated topological order in topological insulators (TIs) and superconductors (TSCs) has been recognized as the principle behind various seemingly unrelated phenomena in these materials\cite{Hasan10,Qi11}. The topological phases of these materials are characterized by topological invariants that are derived from the Bloch state of the electrons or quasiparticles in momentum space, which have been well-understood within the context of symmetry classification that classifies the Dirac Hamiltonians of these materials according to their symmetries and dimensions\cite{Schnyder08,Ryu10,Kitaev09,Chiu16}. On the other hand, it has been pointed out that several kinds of topological invariants can be expressed as real space quantities completely defined from lattice eigenstates, giving rise to the notion of topological markers. The earliest and most widely investigated example is the Chern marker in 2D time-reversal (TR) symmetry-breaking systems, where the corresponding topological invariant is the Chern number calculated from the momentum-integration of Berry curvature\cite{Bianco11,Prodan10,Prodan10_2,Prodan11}. Through rewriting the Berry curvature into projectors to the valence and conduction bands, which can further be expressed in terms of projectors to the filled and empty states of the lattice Hamiltonian after a momentum integration, the diagonal element of the resulting Chern operator on lattice site ${\bf r}$ yields the correct Chern number\cite{Bianco11}. Since the discovery of Chern marker, various topological markers have been proposed to generalize this concept to other dimensions and symmetry classes\cite{Loring10,Bianco13,MondragonShem14,Marrazzo17,Cardano17,Meier18,Huang18,Huang18_2,Focassio21,Sykes21,Jezequel22,Wang22,Hannukainen22}, which have been proved to be a powerful tool to investigate how the real space inhomogeneity, such as disorder and interfaces, can influence the topological order locally and globally. In addition, several theoretical proposals suggest that some topological markers may be measured by real space experiments\cite{Molignini22_Chern_marker,Chen22_spin_Chern_marker}.

In this paper, we address two important issues that naturally arises along the development of topological marker theories: (1) Firstly, is it possible to formulate a universal topological marker that can be ubiquitously applied to lattice models of TIs and TSCs in any dimension and symmetry class? This question is raised because recently, a wrapping number has been proposed as the universal momentum space topological invariant in any dimension and symmetry class\cite{vonGersdorff21_unification}, which has the physical meaning as the number of times that the Brillouin zone (BZ) torus $T^{D}$ wraps around the target sphere $S^{D}$ of the Dirac Hamiltonian. Similar to the derivation of Chern marker from rewriting the Chern number into real space\cite{Bianco11}, we demonstrate that the wrapping number can always be expressed in real space as the trace of an object that we call the topological operator. The diagonal element of the topological operator at lattice site ${\bf r}$ then corresponds in a universal topological marker. (2) Secondly, can topological phase transitions (TPTs) also be detected in real space by some universal quantity valid for any dimension and symmetry class? This question arises because the integrand of the wrapping number, which plays the role of the Jacobian of the aforementioned $T^{D}\rightarrow S^{D}$ map, has a universal critical behavior in Dirac models, namely it narrows and flips sign at the gap-closing high symmetry point ${\bf k}_{0}$ as the system crosses TPTs\cite{Chen16,Chen16_2,Chen17,Chen19_AMS_review,Chen19_universality}. We show that this critical behavior can be detected ubiquitously by the $({\bf r+R,r})$-th off-diagonal element of the topological operator that we call the nonlocal topological marker, which is equivalently the Fourier transform of the Jacobian that narrows and flips sign, and thus decays in real space with a decay length that diverges at TPTs.

The structure of the paper is organized in the following manner. In Sec.~II, we introduce the formalism that rewrites the wrapping number into projectors to valence and conduction band states, and how it can further be expressed in terms of filled and empty lattice eigenstates, yielding the topological operator. The local and nonlocal topological markers are further introduced as the diagonal and off-diagonal elements of the topological operator, respectively, and their interpretations in terms of Wannier states are given. In Sec.~III, IV, and V, we explicitly construct the topological operator for the topologically nontrivial symmetry classes in 3D, 2D, and 1D, and examine various prototype lattice models to demonstrate the universal features of the local and nonlocal topological markers. Section VI summarizes the results, and lists a numbers of open questions that remain to be explored.

\section{General formalism in any dimension and symmetry class \label{sec:general_formalism}}

\subsection{Topological operators \label{sec:topological_operators}}

Our aim is to formulate a real space topological marker for TIs and TSCs in $D$-dimension described by Dirac Hamiltonian $H={\bf d}({\bf k})\cdot{\boldsymbol\Gamma}$, where $\Gamma_{i}=(\Gamma_{0},\Gamma_{1}...\Gamma_{2n})$ are the $n$-th order Dirac matrices of dimension $2^{n}\times 2^{n}$ that satisfy $\left\{\Gamma_{i},\Gamma_{j}\right\}=2\delta_{ij}$, and ${\bf d}({\bf k})=(d_{0},d_{1}...d_{D})$ characterizes the momentum dependence of the Hamiltonian\cite{Schnyder08,Ryu10,Chiu16}. It is often more convenient to work on the spectrally flattened Dirac Hamiltonian $\tilde{Q}={\bf n}({\bf k})\cdot{\boldsymbol\Gamma}$ at momentum ${\bf k}$, where ${\bf n}={\bf d}/|{\bf d}|$ is a momentum-dependent unit vector. The precise form of the $\Gamma$-matrices depends on the dimension and symmetry class of the system at hand. Nevertheless, it has recently been discovered that all the dimensions and symmetry classes can be described by a universal topological invariant calculated from momentum-integration of the cyclic derivative of the ${\bf d}$-vector or ${\bf n}$-vector
\begin{eqnarray}
{\rm deg}[{\bf n}]&=&\frac{1}{V_{D}}\int d^{D}{\bf k}\,\varepsilon_{i_{0}...i_{D}}\frac{1}{|{\bf d}|^{D+1}}d^{i_{0}}\partial_{1}d^{i_{1}}...\partial_{D}d^{i_{D}}
\nonumber \\
&=&\frac{1}{V_{D}}\int d^{D}{\bf k}\,\varepsilon_{i_{0}...i_{D}}n^{i_{0}}\partial_{1}n^{i_{1}}...\partial_{D}n^{i_{D}},
\label{wrapping_number_formula}
\end{eqnarray}
which has been referred to as the wrapping number or degree of the map that counts the number of times the $T^{D}$ BZ wraps around the unit sphere $S^{D}$ that the ${\bf n}$-vector forms, and the integrand $J_{\bf k}=\varepsilon_{i_{0}...i_{D}}n^{i_{0}}\partial_{1}n^{i_{1}}...\partial_{D}n^{i_{D}}$ is the Jacobian of the map\cite{vonGersdorff21_unification}. Here $V_{D}=2\pi^{(D+1)/2}/\Gamma(\frac{D+1}{2})$ is the volume of the $D$-sphere of unit radius, and $\partial_{j}\equiv\partial/\partial k_{j}$. The true topological invariant is either ${\rm deg}[{\bf n}]$ if the system belongs to the so-called primary or complex series, $2\,{\rm deg}[{\bf n}]$ for the even series, and $(-1)^{{\rm deg}[{\bf n}]}$ for the first and second descendants.  

%{\cblue (Make a schematic figure of the wrapping from $T^{D}$ to $S^{D}$)}

We now elaborate that the ${\rm deg}[{\bf n}]$ in Eq.~(\ref{wrapping_number_formula}) can be expressed as the momentum integration of the trace ${\rm Tr}[W\tilde{Q}(d\tilde{Q})^{D}]$, where 
\begin{eqnarray}
\tilde{Q}(d\tilde{Q})^{D}\equiv\tilde{Q}\partial_{1}\tilde{Q}\partial_{2}\tilde{Q}...\partial_{D}\tilde{Q}.
\end{eqnarray}
and $W$ is the product of Dirac matrices that are omitted in the Dirac Hamiltonian for the system at hand, or the identity matrix $W=I$ if all the Dirac matrices are used. This connection is made because the trace of the product of all the $n$-th order $\Gamma$-matrices is a constant
\begin{eqnarray}
{\rm Tr}\left[\Gamma_{0}\Gamma_{1}...\Gamma_{2n}\right]=2^{n}c,
\label{Tr_all_Gamma}
\end{eqnarray}
where the prefactor $c=\left\{1,-1,i,-i\right\}$ depends on the representation of the $\Gamma$-matrices for the system at hand. Now suppose for a specific TI or TSC, the Dirac Hamiltonian uses only $\left\{\Gamma_{0},\Gamma_{1},...\Gamma_{D}\right\}$, leaving $\left\{\Gamma_{D+1},\Gamma_{D+2},...\Gamma_{2n}\right\}$ unused. If we define the product of all the unused ones to be $W=\Gamma_{D+1}\Gamma_{D+2}...\Gamma_{2n}$, it then follows that (repeating indices are summed)
\begin{eqnarray}
&&{\rm Tr}[W\tilde{Q}(d\tilde{Q})^{D}]
\nonumber \\
&&={\rm Tr}\left[\Gamma_{D+1}\Gamma_{D+2}...\Gamma_{2n}\Gamma_{i_{0}}\Gamma_{i_{1}}...\Gamma_{i_{D}}\right]
n^{i_{0}}\partial_{1}n^{i_{1}}...\partial_{D}n^{i_{D}}
\nonumber \\
&&=2^{n}c\,\varepsilon_{i_{0}...i_{D}}n^{i_{0}}\partial_{1}n^{i_{1}}...\partial_{D}n^{i_{D}},
\end{eqnarray}
where the permutation $\varepsilon_{i_{0}...i_{D}}$ comes from the fact that the $\Gamma$-matrices anticommute, and the trace is nonzero and given by Eq.~(\ref{Tr_all_Gamma}) only if every $\Gamma$-matrix appears once and only once. Taking a momentum integration and comparing with Eq.~(\ref{wrapping_number_formula}), we obtain
\begin{eqnarray}
{\rm deg}[{\bf n}]=\frac{(2\pi)^{D}}{2^{n}c\,V_{D}}\int\frac{d^{D}{\bf k}}{(2\pi)^{D}}{\rm Tr}[W\tilde{Q}(d\tilde{Q})^{D}].
\label{degn_IntTrWQdQ}
\end{eqnarray}
This quantity $\int d^{D}{\bf k}\,{\rm Tr}[W\tilde{Q}(d\tilde{Q})^{D}]$ is our bridge to a real space topological marker, because it allows to adopt the projector algebra that originally derives the Chern marker\cite{Bianco11}. To see this, we observe that the spectrally flattened Hamiltonian $\tilde{Q}$ can be separated into the projector $p$ into the valence band states $|n({\bf k})\rangle$ and the projector $q$ into the conduction band states $|m({\bf k})\rangle$, 
\begin{eqnarray}
\tilde{Q}=q-p,\;\;\;p=\sum_{n}|n\rangle\langle n|,\;\;\;q=\sum_{m}|m\rangle\langle m|.
\end{eqnarray}
and $q+p=I$. As a result, the derivative of $\tilde{Q}$ over a certain component of momentum $\partial/\partial k_{j}\equiv\partial_{j}$ is equivalently $\partial_{i}\tilde{Q}=2\partial_{i}q=-2\partial_{i}p$. Consequently, we can write $\tilde{Q}(d\tilde{Q})^{D}$ into the form that consists of alternating derivatives of $p$ and $q$.

Consider first $D=$ odd, in which case the integrand of Eq.~(\ref{degn_IntTrWQdQ}) reads
\begin{eqnarray}
&&W\tilde{Q}(d\tilde{Q})^{D}|_{D\in odd}=2^{D}(-1)^{(D+1)/2}
\nonumber \\
&&\times W\left\{q\,\partial_{1}p\,\partial_{2}q...
\partial_{D}p+p\,\partial_{1}q\,\partial_{2}p...
\partial_{D}q\right\}
\nonumber \\
&&=2^{D}(-1)^{(D+1)/2}\sum_{m_{1}\sim m_{(D+1)/2}}\sum_{n_{1}\sim n_{(D+1)/2}}
\nonumber \\
&&W\left\{|m_{1}\rangle\langle m_{1}|\partial_{i_{1}}n_{1}\rangle\langle n_{1}|\partial_{i_{2}}m_{2}\rangle...\right.
\nonumber \\
&&\left. ...\langle m_{(D+1)/2}|\partial_{i_{D}}n_{(D+1)/2}\rangle\langle n_{(D+1)/2}|+(m\leftrightarrow n)\right\}.
\end{eqnarray}
The motivation to rewrite it into this form of alternating $p$ and $q$ is to use the identity\cite{Bianco11} 
\begin{eqnarray}
\langle m|\partial_{i}|n\rangle=-i\langle\psi_{m}|{\hat i}|\psi_{n}\rangle,
\label{mdin_psiipsi}
\end{eqnarray}
provided $n\neq m$, where the $|\psi_{n}\rangle=|\psi_{n}({\bf k})\rangle$ is the full wave function that satisfies $\langle{\bf r}|\psi_{n}({\bf k})\rangle=\psi_{n{\bf k}}({\bf r})=u_{n{\bf k}}({\bf r})e^{i{\bf k\cdot r}}=\langle{\bf r}|n({\bf k})\rangle e^{i{\bf k\cdot r}}$ with $u_{n{\bf k}}({\bf r})$ the Bloch periodic part of the wave function. Here ${\hat i}$ is the position operator, which is a diagonal matrix where all the $2^{n}$ internal degrees of freedom (spin, orbit, particle-hole, etc) within a unit cell located at the Bravais lattice vector ${\bf r}=(x,y,z...)$ are assigned with the same ${\bf r}$. This identity allows us to write 
\begin{eqnarray}
&&\int\frac{d^{D}{\bf k}}{(2\pi)^{D}}{\rm Tr}[W\tilde{Q}(d\tilde{Q})^{D}]_{D\in odd}
\nonumber \\
&&=2^{D}i\int\frac{d^{D}{\bf k}}{(2\pi)^{D}}\varepsilon^{i_{1}i_{2}...i_{D}}\sum_{m_{1}\sim m_{(D+1)/2}}\sum_{n_{1}\sim n_{(D+1)/2}}
\nonumber \\
&&{\rm Tr}\left\{W|\psi_{m_{1}}\rangle\langle \psi_{m_{1}}|{\hat i_{1}}|\psi_{n_{1}}\rangle\langle \psi_{n_{1}}|...\right.
\nonumber \\
&&\left. ...\langle \psi_{m_{(D+1)/2}}|{\hat i_{D}}|\psi_{n_{(D+1)/2}}\rangle\langle \psi_{n_{(D+1)/2}}|+(m\leftrightarrow n)\right\}
\nonumber \\
&&=2^{D}i{\rm Tr}\left[W\int\frac{d^{D}{\bf k}_{m_{1}}}{(2\pi)^{D}}\sum_{m_{1}}|\psi_{m_{1}}\rangle\langle \psi_{m_{1}}|{\hat i_{1}}\right.
\nonumber \\
&&\times\int\frac{d^{D}{\bf k}_{n_{1}}}{(2\pi)^{D}}\sum_{n_{1}}|\psi_{n_{1}}\rangle\langle \psi_{n_{1}}|{\hat i_{2}}...
\nonumber \\
&&...\int\frac{d^{D}{\bf k}_{m_{(D+1)/2}}}{(2\pi)^{D}}\sum_{m_{(D+1)/2}}|\psi_{m_{(D+1)/2}}\rangle\langle \psi_{m_{(D+1)/2}}|{\hat i_{D}}
\nonumber \\
&&\times\int\frac{d^{D}{\bf k}_{n_{(D+1)/2}}}{(2\pi)^{D}}\sum_{n_{(D+1)/2}}|\psi_{n_{(D+1)/2}}\rangle\langle \psi_{n_{(D+1)/2}}|
\nonumber \\
&&\left.+(n\leftrightarrow m)\right].
\label{TrWQdQ_Dodd_projector}
\end{eqnarray}
In the last step, we identity the projector to the valence bands integrated over momentum as the projector to the filled band states $|E_{n}\rangle$ of a lattice Hamiltonian, and likewisely the projector to the conduction bands integrated over momentum as the projector to the empty band states $|E_{m}\rangle$
\begin{eqnarray}
&&\int\frac{d^{D}{\bf k}_{m_{j}}}{(2\pi)^{D}}\sum_{m_{j}}|\psi_{m_{j}}\rangle\langle \psi_{m_{j}}|=\sum_{m}|E_{m}\rangle\langle E_{m}|\equiv Q,
\nonumber \\
&&\int\frac{d^{D}{\bf k}_{n_{j}}}{(2\pi)^{D}}\sum_{n_{j}}|\psi_{n_{j}}\rangle\langle \psi_{n_{j}}|=\sum_{n}|E_{n}\rangle\langle E_{n}|\equiv P,
\label{projector_lattice_eigenstates}
\end{eqnarray}
which are $\left(L^{D}2^{n}\right)\times \left(L^{D}2^{n}\right)$ matrices, where $L^{D}$ is the total number of unit cells in the lattice, and each unit cell contains $2^{n}$ degrees of freedom. Equation (\ref{TrWQdQ_Dodd_projector}) then becomes
\begin{eqnarray}
&&\int\frac{d^{D}{\bf k}}{(2\pi)^{D}}{\rm Tr}[W\tilde{Q}(d\tilde{Q})^{D}]_{D\in odd}
\nonumber \\
&&=2^{D}i{\rm Tr}\left[WQ\,{\hat i_{1}}P\,{\hat i_{2}}...Q\,{\hat i_{D}}P+WP\,{\hat i_{1}}Q\,{\hat i_{2}}...P\,{\hat i_{D}}Q\right],
\nonumber\\
\label{TrWQdQ_Dodd_TrQdP}
\end{eqnarray}
where in the second line we have enlarged $W\rightarrow W\otimes I_{L^{D}\times L^{D}}$. In this way we have written the momentum space topological invariant into a object that is completely defined from the eigenstates of a lattice Hamiltonian.

In $D=$ even dimensions, the construction is similar. We again seek to write the integrand $W\tilde{Q}(d\tilde{Q})^{D}$ into alternating derivatives of $p$ and $q$, which in even dimensions becomes
\begin{eqnarray}
&&W\tilde{Q}(d\tilde{Q})^{D}|_{D\in even}=2^{D}(-1)^{D/2}
\nonumber \\
&&\times W\left\{q\,\partial_{i_{1}}p\,\partial_{i_{2}}q...
\partial_{i_{D}}q-p\,\partial_{i_{1}}q\,\partial_{i_{2}}p...
\partial_{i_{D}}p\right\}
\nonumber \\
&&=2^{D}(-1)^{D/2}\sum_{m_{1}\sim m_{D/2+1}}\sum_{n_{1}\sim n_{D/2}}
\nonumber \\
&&W\left\{|m_{1}\rangle\langle m_{1}|\partial_{i_{1}}n_{1}\rangle\langle n_{1}|\partial_{i_{2}}m_{2}\rangle...\right.
\nonumber \\
&&\left. ...\langle n_{D/2}|\partial_{i_{D}}m_{D/2+1}\rangle\langle m_{D/2+1}|-(m\leftrightarrow n)\right\},
\end{eqnarray}
whose momentum integration becomes
\begin{eqnarray}
&&\int\frac{d^{D}{\bf k}}{(2\pi)^{D}}{\rm Tr}[W\tilde{Q}(d\tilde{Q})^{D}]_{D\in even}
\nonumber \\
&&=2^{D}\int\frac{d^{D}{\bf k}}{(2\pi)^{D}}\sum_{m_{1}\sim m_{D/2+1}}\sum_{n_{1}\sim n_{D/2}}
\nonumber \\
&&{\rm Tr}\left\{W|\psi_{m_{1}}\rangle\langle \psi_{m_{1}}|{\hat i_{1}}|\psi_{n_{1}}\rangle\langle \psi_{n_{1}}|...\right.
\nonumber \\
&&\left. ...\langle \psi_{n_{D/2}}|{\hat i_{D}}|\psi_{m_{D/2+1}}\rangle\langle \psi_{m_{D/2+1}}|-(m\leftrightarrow n)\right\}
\nonumber \\
&&=2^{D}{\rm Tr}\left[W\int\frac{d^{D}{\bf k}_{m_{1}}}{(2\pi)^{D}}\sum_{m_{1}}|\psi_{m_{1}}\rangle\langle \psi_{m_{1}}|{\hat i_{1}}\right.
\nonumber \\
&&\times\int\frac{d^{D}{\bf k}_{n_{1}}}{(2\pi)^{D}}\sum_{n_{1}}|\psi_{n_{1}}\rangle\langle \psi_{n_{1}}|{\hat i_{2}}...
\nonumber \\
&&...\int\frac{d^{D}{\bf k}_{n_{D/2}}}{(2\pi)^{D}}\sum_{n_{D/2}}|\psi_{n_{D/2}}\rangle\langle \psi_{n_{D/2}}|{\hat i_{D}}
\nonumber \\
&&\times\int\frac{d^{D}{\bf k}_{m_{D/2+1}}}{(2\pi)^{D}}\sum_{m_{D/2+1}}|\psi_{m_{D/2+1}}\rangle\langle \psi_{m_{D/2+1}}|
\nonumber \\
&&-\left.(n\leftrightarrow m)\right].
\label{TrWQdQ_Deven_projector}
\end{eqnarray}
Using the projectors in Eq.~(\ref{projector_lattice_eigenstates}), we arrive at 
\begin{eqnarray}
&&\int\frac{d^{D}{\bf k}}{(2\pi)^{D}}{\rm Tr}[W\tilde{Q}(d\tilde{Q})^{D}]_{D\in even}
\nonumber \\
&&=2^{D}{\rm Tr}\left[WQ\,{\hat i_{1}}P\,{\hat i_{2}}...P\,{\hat i_{D}}Q-WP\,{\hat i_{1}}Q\,{\hat i_{2}}...Q\,{\hat i_{D}}P\right].
\nonumber \\
\label{TrWQdQ_Deven_TrQdP}
\end{eqnarray}
Once again we have written the topological invariant into a form that is completely defined from the lattice eigenstates.

Equations (\ref{TrWQdQ_Dodd_TrQdP}) and (\ref{TrWQdQ_Deven_TrQdP}) suggest a universal topological operator of the form
\begin{eqnarray}
{\hat {\cal C}}=N_{D}W\left[Q\,{\hat i_{1}}P\,{\hat i_{2}}...\,{\hat i_{D}}{\cal O}+(-1)^{D+1}P\,{\hat i_{1}}Q\,{\hat i_{2}}...{\hat i_{D}}{\overline{\cal O}}\right],
\nonumber \\
\label{topological_operator}
\end{eqnarray}
where the last operators $\left\{{\cal O},\overline{\cal O}\right\}=\left\{P,Q\right\}$ if $D=$ odd, and $\left\{{\cal O},\overline{\cal O}\right\}=\left\{Q,P\right\}$ if $D=$ even owing to the alternating order of the projectors $Q$ and $P$. Using Eqs.~(\ref{degn_IntTrWQdQ}), (\ref{TrWQdQ_Dodd_TrQdP}), and (\ref{TrWQdQ_Deven_TrQdP}), one sees that the wrapping number in Eq.~(\ref{wrapping_number_formula}) is equal to the trace of this operator 
\begin{eqnarray}
{\rm deg}[{\bf n}]=\frac{1}{L^{D}}{\rm Tr}[\,{\hat{\cal C}}\,].
\label{degn_TrC_correspondence}
\end{eqnarray}
where $L^{D}$ is the total number of unit cells, and ${\rm Tr}[...]$ represents the trace over all the lattice sites. The normalization factor $N_{D}$ in Eq.~(\ref{topological_operator}) has the expression
\begin{eqnarray}
N_{D}=\frac{i^{D}2^{2D-n}\pi^{D}}{c\,V_{D}},
\label{ND_factor}
\end{eqnarray}
which depends on the dimension $D$, the volume $V_{D}=\left\{V_{1},V_{2},V_{3}...\right\}=\left\{2\pi,4\pi,2\pi^{2}...\right\}$, the order $n$ and the prefactor $c={\rm Tr}\left[\Gamma_{0}\Gamma_{1}...\Gamma_{2n}\right]/2^{n}=\left\{1,-1,i,-i\right\}$ of the representation of $\Gamma$-matrices for the system under question. Equations (\ref{topological_operator}) to (\ref{ND_factor}) are the central results of this work, and we will demonstrate their validity using concrete models in the following sections.

%is chosen such that Eq.~(\ref{degn_TrC_correspondence}) is satisfied. In practice, the real or imaginary $N_{D}$ may be best determined directly by numerical calculations. {\cblue (Can I unambiguously figure out the normalization factor $N_{D}$? Does it depend on the precise form of $W$?)} 

%It can be easily verified that the expression in the even dimension goes back to the well-known Chern marker by setting $D=2$ and $W=I$, and the spin Chern marker by setting $D=2$ and using the spin operator $W=\sigma^{z}$. {\cblue (cite my papers)} 

\subsection{Local and nonlocal topological markers}

Similar to the original construction of Chern marker as the diagonal elements of Chern operator\cite{Bianco11}, the correspondence between the wrapping number and the trace of the topological operator in Eq.~(\ref{degn_TrC_correspondence}) suggests to define the (local) topological marker on a lattice site ${\bf r}$ by
\begin{eqnarray}
{\cal C}({\bf r})=\langle{\bf r}|{\hat{\cal C}}|{\bf r}\rangle=\sum_{\sigma}\langle{\bf r}\sigma|{\hat{\cal C}}|{\bf r}\sigma\rangle,
\end{eqnarray}
i.e., the marker is the diagonal element of topological operator at ${\bf r}$. Here $\sum_{\sigma}$ represents the summation over all the $2^{n}$ internal degrees of freedom inside the unit cell at ${\bf r}$, such as spin, orbital, particle-hole, etc.

\begin{figure}[ht]
\begin{center}
\includegraphics[clip=true,width=0.9\columnwidth]{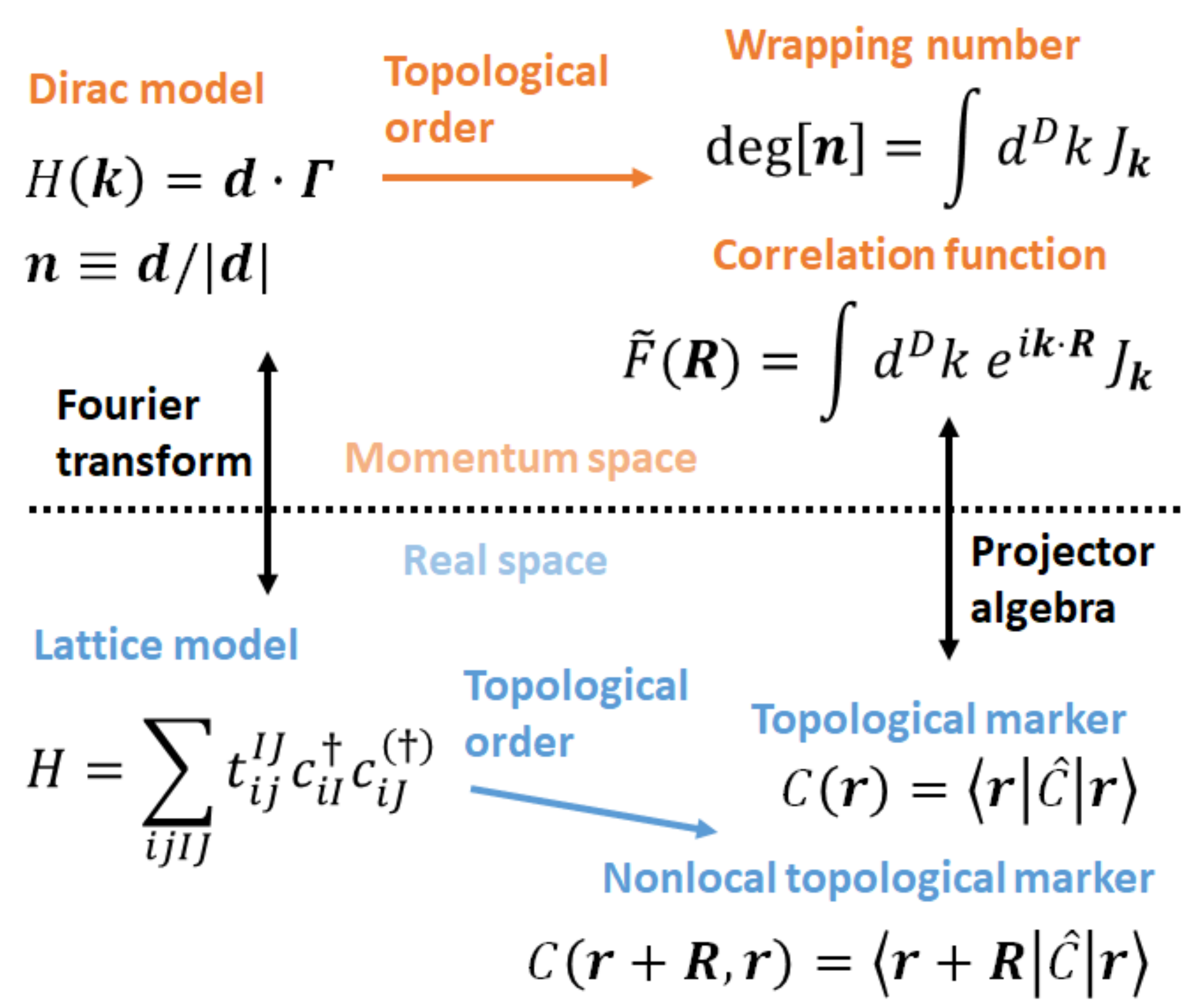}
\caption{Summary of the correspondence between the topological order in momentum space calculated from the Jacobian $J_{\bf k}=\varepsilon_{i_{0}...i_{D}}n^{i_{0}}\partial_{1}n^{i_{1}}...\partial_{D}n^{i_{D}}$, where $n^{i}$ characterizes the spectrally flattened Dirac Hamiltonian $\tilde{Q}={\bf n}\cdot{\boldsymbol\Gamma}$, and that in real space calculated from the topological operator ${\cal C}$ constructed from projectors and position operators. For homogeneous systems in the thermodynamic limit, the wrapping number is equal to the topological marker ${\rm deg}[{\bf n}]={\cal C}({\bf r})$, and the Wannier state correlation function is equivalently the nonlocal topological marker $\tilde{F}({\bf R})={\cal C}({\bf r+R,r})$. } 
\label{fig:kspace_realspace_correspondence}
\end{center}
\end{figure}

We further introduce a nonlocal topological marker as the $({\bf r+R,r})$-th off-diagonal matrix element of the topological operator\cite{Molignini22_Chern_marker,Chen22_spin_Chern_marker}
\begin{eqnarray}
{\cal C}({\bf r+R,r})=\langle{\bf r+R}|{\hat{\cal C}}|{\bf r}\rangle,
\end{eqnarray}
where ${\bf R}$ is a Bravais lattice vector. For a homogeneous lattice model in the thermodynamic limit, this nonlocal marker is independent of ${\bf r}$, and is equivalently the Fourier transform of the integrand of the wrapping number in Eq.~(\ref{wrapping_number_formula})
\begin{eqnarray}
&&{\cal C}({\bf r+R,r})=\tilde{F}({\bf R})
\nonumber \\
&&\equiv\frac{1}{V_{D}}\int d^{D}{\bf k}\,\varepsilon_{i_{0}...i_{D}}n^{i_{0}}\partial_{1}n^{i_{1}}...\partial_{D}n^{i_{D}}e^{i{\bf k\cdot R}}.
\end{eqnarray}
that has been previously denoted by $\tilde{F}({\bf R})$\cite{Chen17,Chen19_universality,Chen19_AMS_review}, and has the physical meaning as a correlation function that measures the overlap between Wannier states that are a distance ${\bf R}$ apart, as we shall see in Sec.~\ref{sec:Wannier_representation}. The identification ${\cal C}({\bf r+R,r})=\tilde{F}({\bf R})$ can be seen by considering equivalently the Fourer transform of ${\rm Tr}[W\tilde{Q}(d\tilde{Q})^{D}]$, which contains the projection $\langle{\bf r}|W|\psi_{m_{1}}\rangle e^{i{\bf k\cdot R}}$ of the first ket state of Eqs.~(\ref{TrWQdQ_Dodd_projector}) and (\ref{TrWQdQ_Deven_projector}) that is equal to $\langle{\bf r+R}|W|\psi_{m_{1}}\rangle$
\begin{eqnarray}
&&\langle{\bf r}|W|\psi_{m_{1}}\rangle e^{i{\bf k\cdot R}}=e^{i{\bf k(r+R)}}W\,u_{m_{1}{\bf k}}({\bf r+R})
\nonumber \\
&&=W\,\psi_{m_{1}{\bf k}}({\bf r+R})=\langle{\bf r+R}|W|\psi_{m_{1}}\rangle,
\end{eqnarray}
owing to the cell periodicity of the Bloch state $u_{m_{1}{\bf k}}({\bf r+R})=u_{m_{1}{\bf k}}({\bf r})$. The correspondence between the momentum space topological invariants and the real space topological markers is summarized schematically in Fig.~\ref{fig:kspace_realspace_correspondence}.

The spatial profile of the nonlocal marker ${\cal C}({\bf r+R,r})$ decays with ${\bf R}$, with a decay length $\xi$ that diverges at TPTs, thereby serving as a faithful quantity to identify TPTs. This can be seen by considering the linear Dirac model $d^{0}=M$, $d^{i\neq 0}=vk_{i}$, which is a low energy effective model near the gap-closing momentum that describes the majority of TPTs. The integrand of Eq.~(\ref{wrapping_number_formula}) has a Lorentzian shape in this model
\begin{eqnarray}
&&\varepsilon_{i_{0}...i_{D}}\frac{1}{|{\bf d}|^{D+1}}d^{i_{0}}\partial_{1}d^{i_{1}}...\partial_{D}d^{i_{D}}=\frac{M}{\left[M^{2}+v^{2}k^{2}\right]^{(D+1)/2}}
\nonumber \\
&&\approx\frac{{\rm Sgn}(M)|M|^{-D}}{1+\xi^{2}k^{2}},
\label{FkM_Lorentzian}
\end{eqnarray}
implying that its Fourier transform ${\cal C}({\bf r+R,r})$ decays with a correlation length $\xi\sim|M|^{-\nu}$ that diverges at the critical point $M_{c}=0$, with the critical exponent $\nu=1$. In addition, owing to the relation between the integrand and the quantum metric of the valence band state\cite{Provost80}, a relation that has been called the metric-curvature correspondence\cite{vonGersdorff21_metric_curvature}, the integrand at ${\bf k}=0$ that scales like $\sim|M|^{-D}$ has the meaning as the fidelity susceptibility near TPTs\cite{Panahiyan20_fidelity}, and hence has been assigned with the exponent $\gamma=D$.

\subsection{Wannier state representations \label{sec:Wannier_representation}}

We proceed to elaborate that in the homogeneous and thermodynamic limit, both the local and nonlocal topological markers can be expressed in terms of overlap of Wannier states. Given the Bloch states $|\ell\rangle=|\ell_{\bf k}\rangle$ of either the valence $\ell=n$ or conduction $\ell=m$ band states, we can introduce the Wannier state $|{\bf R}\ell\rangle$ by
\begin{eqnarray}
|\ell_{{\bf k}}\rangle=\sum_{{\bf R}}e^{-i {\bf k}\cdot({\hat{\bf r}}-{\bf R})}|{\bf R}\ell\rangle,\;\;\;
|{\bf R} \ell\rangle=\sum_{\bf k}e^{i {\bf k}\cdot({\hat{\bf r}}-{\bf R})}|\ell_{{\bf k}}\rangle.\;\;\;\;\;\;\;
\label{Wannier_basis}
\end{eqnarray}
Inserting these definitions into Eqs.~(\ref{TrWQdQ_Dodd_projector}) and (\ref{TrWQdQ_Deven_projector}) yields an expression for the local topological marker
\begin{eqnarray}
&&{\cal C}({\bf r})|_{D\in odd}\propto\int\frac{d^{D}{\bf k}}{(2\pi)^{D}}{\rm Tr}[W\tilde{Q}(d\tilde{Q})^{D}]_{D\in odd}
\nonumber \\
&&=2^{D}i\varepsilon^{i_{1}...i_{D}}\sum_{R_{1}\sim R_{2D+1}}\sum_{m_{1}\sim m_{(D+1)/2}}\sum_{n_{1}\sim n_{(D+1)/2}}
\nonumber \\
&&\langle{\bf R}_{1}n_{(D+1)/2}|W|{\bf R}_{2}m_{1}\rangle\langle{\bf R}_{3}m_{1}|{\hat i}_{1}|{\bf R}_{4}n_{1}\rangle...
\nonumber \\
&&\langle{\bf R}_{2D+1}m_{(D+1)/2}|{\hat i}_{D}|({\bf R}_{1}-{\bf R}_{2}+...+{\bf R}_{2D+1})n_{(D+1)/2}\rangle
\nonumber \\
&&+(n\leftrightarrow m),
\nonumber \\
&&{\cal C}({\bf r})|_{D\in even}\propto\int\frac{d^{D}{\bf k}}{(2\pi)^{D}}{\rm Tr}[W\tilde{Q}(d\tilde{Q})^{D}]_{D\in even}
\nonumber \\
&&=2^{D}\varepsilon^{i_{1}...i_{D}}\sum_{R_{1}\sim R_{2D+1}}\sum_{m_{1}\sim m_{D/2+1}}\sum_{n_{1}\sim n_{D/2}}
\nonumber \\
&&\langle{\bf R}_{1}m_{D/2+1}|W|{\bf R}_{2}m_{1}\rangle\langle{\bf R}_{3}m_{1}|{\hat i}_{1}|{\bf R}_{4}n_{1}\rangle...
\nonumber \\
&&\langle{\bf R}_{2D+1}n_{D/2}|{\hat i}_{D}|({\bf R}_{1}-{\bf R}_{2}+{\bf R}_{3}...+{\bf R}_{2D+1})m_{D/2+1}\rangle
\nonumber \\
&&-(n\leftrightarrow m).
\label{Cr_Wannier_representation}
\end{eqnarray}
Likewisely, the nonlocal topological markers ${\cal C}({\bf r+R,r})$ in the homogeneous limit in either even or odd dimensions can also be expressed in terms of Wannier states, which is simply given by the results in Eq.~(\ref{Cr_Wannier_representation}) with the last position argument replaced by $({\bf R}_{1}-{\bf R}_{2}+{\bf R}_{3}...+{\bf R}_{2D+1})\rightarrow({\bf R}_{1}-{\bf R}_{2}+{\bf R}_{3}...+{\bf R}_{2D+1}-{\bf R})$. As a result, the nonlocal marker ${\cal C}({\bf r+R,r})$ has the physical meaning as the measure of the overlap of Wannier states weighted by the position operators, which decays with ${\bf R}$ according to the argument after Eq.~(\ref{FkM_Lorentzian}). Moreover, the decay length $\xi$ diverges at TPTs, and hence ${\cal C}({\bf r+R,r})$ serves as a faithful correlator that characterizes the quantum criticality near TPTs.

%{\cblue (2) Say that in this paper we focus on cubic lattice models, whereas other types of lattice structures, like Haldane model and Kane-Mele model that are defined on a honeycomb lattice, will be addressed in the future. }

\subsection{Applications to lattice models from 1D to 3D}

The TIs and TSCs can be classified according to the TR, particle-hole (PH) and chiral symmetries of the single particle Hamiltonian defined by\cite{Schnyder08,Ryu10,Kitaev09,Chiu16}
\begin{eqnarray}
&&TH({\bf k})T^{-1}=H(-{\bf k}),
\nonumber \\
&&CH({\bf k})C^{-1}=-H(-{\bf k}),
\nonumber \\
&&SH({\bf k})S^{-1}=-H({\bf k}),
\label{General_symmetries_TR_PH_CH}
\end{eqnarray}
yielding a total of 10 symmetry classes. The result of the classification gives 5 topologically nontrivial symmetry classes in each spatial dimension $D$. For practical reasons, in the following sections, we explicitly construct the topological operators in Eq.~(\ref{topological_operator}) for all the 15 nontrivial symmetry classes from 1D to 3D. Moreover, for those classes described by $2\times 2$ and $4\times 4$ Dirac matrices, which cover 13 out of the 15 nontrivial classes, we will use cubic lattice models to explicitly demonstrate the validity of the local and nonlocal topological markers. The two cases left unexamined are classes CI and CII in 3D described by $8\times 8$ Dirac matrices, which are less explored in the literature and will be left for future investigations.

\begin{figure}[ht]
\begin{center}
\includegraphics[clip=true,width=0.99\columnwidth]{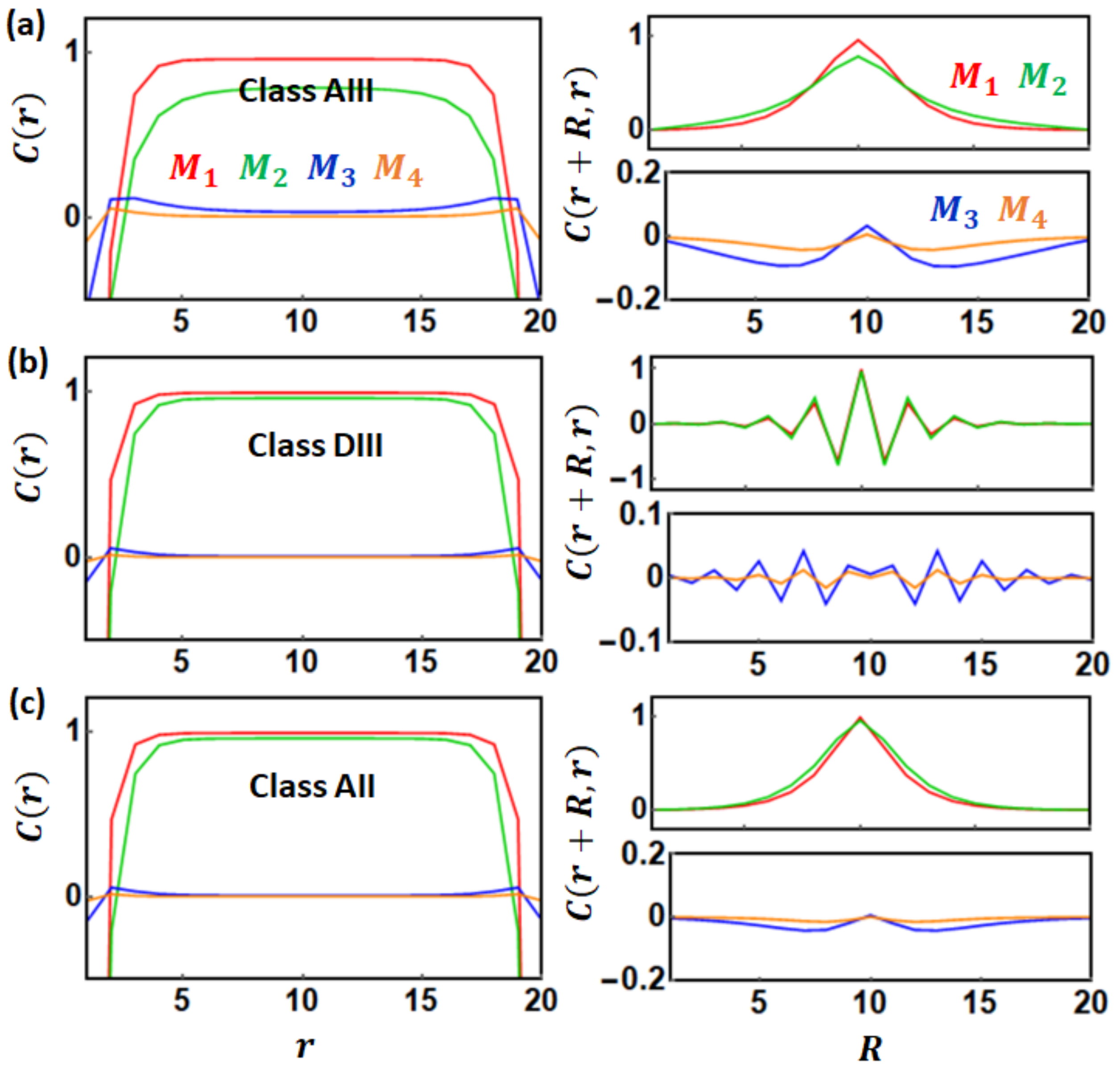}
\caption{Local (left column) and nonlocal (right column) topological markers for lattice models in three out of the five topologically nontrivial symmetry classes in 3D, including (a) a regularized lattice model for class AIII, (b) a lattice model of $^{3}$He B-phase in class DIII, and (c) the prototype 3D TR-symmetric TIs in class AII. } 
\label{fig:3D_results}
\end{center}
\end{figure}

To elaborate the ubiquity of the topological operators and markers, in each lattice model, we choose periodic boundary condition in all spatial directions, focus on one specific critical point of the mass term $M_{c}$, and examine four parameters of $M$ denoted by
\begin{eqnarray}
&&M_{1}:\;\;\;{\rm nontrivial\;phase\;far\;from\;}M_{c}\;({\rm red})
\nonumber \\
&&M_{2}:\;\;\;{\rm nontrivial\;phase\;close\;to\;}M_{c}\;({\rm green})
\nonumber \\
&&M_{3}:\;\;\;{\rm trivial\;phase\;close\;to\;}M_{c}\;({\rm blue})
\nonumber \\
&&M_{4}:\;\;\;{\rm trivial\;phase\;far\;from\;}M_{c}\;({\rm orange})
\label{M1M2M3M4}
\end{eqnarray}
The purpose of examining these 4 parameters is to elaborate that the behavior of local and nonlocal topological markers is the same in any dimension and symmetry class (see Fig.~\ref{fig:3D_results}, \ref{fig:2D_results}, and \ref{fig:1D_results} in the following sections, with the same color code indicated in Eq.~(\ref{M1M2M3M4})): Deep inside the bulk, $M_{1}$ and $M_{2}$ have the same integer-valued topological marker ${\cal C}({\bf r})$ since they are in the same topological phase, but the nonlocal topological marker ${\cal C}({\bf r+R, r})$ of $M_{2}$ has a longer decay length than $M_{1}$ since it is closer to the critical point according to the discussion after Eq.~(\ref{FkM_Lorentzian}). Likewisely, $M_{3}$ and $M_{4}$ have the same ${\cal C}({\bf r})$ deep inside the bulk, but $M_{3}$ has a longer decay length of ${\cal C}({\bf r+R, r})$ than $M_{4}$ since it is closer to the critical point. We also remark that numerically, we find that the local marker at the boundary sites deviates from the bulk value even if periodic boundary condition is employed. This is a feature well-known for this type of construction\cite{Bianco11}, since the position operators $\hat{i}$ in Eq.~(\ref{topological_operator}) do not respect translational invariance. Such an anomaly may be fixed by exponentiating the position operator\cite{Prodan10,Prodan11}, which shall be explored elsewhere.

%{\cblue (1) Note that for the primary series the topological invariant is ${\rm deg}[{\bf n}]={\cal C}({\bf r})$. For the first and second descendents the invariant is $(-1)^{{\rm deg}[{\bf n}]}=(-1)^{{\cal C}({\bf r})}$. For the even series it is $2{\rm deg}[{\bf n}]=2{\cal C}({\bf r})$. So for the even series we should also only get an integer ${\cal C}({\bf r})$ but just multiplied by a factor of 2 at the end. -----Note that I already mentioned this in the beginning of Sec.~II A. }

%The lattice Hamiltonian is constructed from a Fourier transform of the momentum space single-particle Hamiltonian in the usual way
%\begin{eqnarray}
%H=\sum_{IJ{\bf k}}c_{{\bf k}I}^{\dag}H_{IJ}({\bf k})c_{{\bf k}J}
%=\sum_{IJij}c_{iI}^{\dag}H_{ij}^{IJ}c_{jJ},
%\label{Hk_to_Hij}
%\end{eqnarray}
%where $\left\{I,J\right\}$ denote the internal degrees of freedom such as spin, orbital, particle-hole, etc, and $\left\{i,j\right\}$ denote lattice sites on a $D$-dimensional lattice. 

\section{Topological markers in three dimensions}

In 3D TIs and TSCs, the topological operator has the following general form
\begin{eqnarray}
{\hat{\cal C}}_{3D}=N_{D}W
\left[Q\hat{x}P\hat{y}Q\hat{z}P+P\hat{x}Q\hat{y}P\hat{z}Q\right].
\end{eqnarray}
The $W$ matrix has different interpretations in different classes. The numerical results for classes AIII, DIII, and AII, are presented in Fig.~\ref{fig:3D_results} using prototype cubic lattice models simulated on a 3D lattice of dimension $L_{x}\times L_{y}\times L_{z}=20\times 8\times 8$. We use a lattice that is elongated in the ${\hat{\bf x}}$-direction and plot the results along this direction for the sake of increasing numerical accuracy. The details of each symmetry class is described below.

\subsection{3D class AIII \label{sec:3D_class_AIII}}

In 3D classes AIII, the $\Gamma$-matrices are given by\cite{Ryu10} 
\begin{eqnarray}
&&\Gamma_{1\sim 5}=\left(\alpha_{x},\alpha_{y},\alpha_{z},\beta,-i\beta\gamma^{5}\right),
\nonumber \\
&&\alpha_{i}=\left(
\begin{array}{cc}
0 & \sigma_{i} \\
\sigma_{i} & 0
\end{array}
\right),\;\;
\beta=\left(
\begin{array}{cc}
1 & 0 \\
0 & -1
\end{array}
\right),\;\;
\gamma^{5}=\left(
\begin{array}{cc}
0 & 1 \\
1 & 0
\end{array}
\right).\;\;\;
\label{3D_class_AIII_gamma_matrices}
\end{eqnarray}
The chiral operator $S=\beta$ demands $d^{4}({\bf k})=0$, so for a linear Dirac model one chooses $d_{1}=Ak_{x}$, $d_{2}=Ak_{y}$, $d_{3}=Ak_{z}$, and $d_{5}=M+B\sum_{\ell=1}^{3}k_{\ell}^{2}$ as the mass term that contains a quadratic term to avoid fermion doubling. The spinor of 3D class AIII contains only annihilation operators, which we name them generically as $\psi=(c_{{\bf k}1},c_{{\bf k}2},c_{{\bf k}3},c_{{\bf k}4})$, where $c_{{\bf k}J}$ is the $J$-th degree of freedom. Regularizing the linear Dirac model in the whole BZ by
\begin{eqnarray}
&&A\,k_{\ell}\rightarrow A\sin k_{\ell},
\nonumber \\
&&M\rightarrow M+B\sum_{\ell=1}^{D}k_{\ell}^{2}
\rightarrow M+2DB-2B\sum_{\ell=1}^{D}\cos k_{\ell},\;\;\;
\label{k_to_sink}
\end{eqnarray}
with $D=3$ in this case, we may further construct a cubic lattice Hamiltonian by performing a Fourier transform
\begin{eqnarray}
&&\sum_{k}\cos k_{\ell}c_{{\bf k}I}^{\dag}c_{{\bf k}J}\rightarrow\frac{1}{2}
\sum_{i}\left\{c_{iI}^{\dag}c_{i+\ell J}+c_{i+\ell I}^{\dag}c_{iJ}\right\},
\nonumber \\
&&\sum_{k}i\sin k_{\ell}c_{{\bf k}I}^{\dag}c_{{\bf k}J}\rightarrow\frac{1}{2}
\sum_{i}\left\{c_{iI}^{\dag}c_{i+\ell J}-c_{i+\ell I}^{\dag}c_{iJ}\right\}.
\label{regularization_k_to_site}
\end{eqnarray}
Denoting $t=A/2$ and $t'=B$, the resulting lattice model is
\begin{eqnarray}
&&H=\sum_{i}t\left\{-ic_{i1}^{\dag}c_{i+x4}+ic_{i+x1}^{\dag}c_{i4}-ic_{i2}^{\dag}c_{i+x3}+ic_{i+x2}^{\dag}c_{i3}\right\}
\nonumber \\
&&+\sum_{i}t\left\{-c_{i1}^{\dag}c_{i+y4}+c_{i+y1}^{\dag}c_{i4}+c_{i2}^{\dag}c_{i+y3}-c_{i+y2}^{\dag}c_{i3}\right\}
\nonumber \\
&&+\sum_{i}t\left\{-ic_{i1}^{\dag}c_{i+z3}+ic_{i+z1}^{\dag}c_{i3}+ic_{i2}^{\dag}c_{i+z4}-ic_{i+z2}^{\dag}c_{i4}\right\}
\nonumber \\
&&+\sum_{i\delta}it'\left\{c_{i1}^{\dag}c_{i+\delta 3}+c_{i+\delta 1}^{\dag}c_{i3}+c_{i2}^{\dag}c_{i+\delta 4}+c_{i+\delta 2}^{\dag}c_{i4}\right\}
\nonumber \\
&&+\sum_{i}(-iM-i6t')\left\{c_{i1}^{\dag}c_{i3}+c_{i2}^{\dag}c_{i4}\right\}+h.c.
\end{eqnarray}
The omitted Dirac matrix is the chiral operator $W=\beta=\Gamma_{4}=S$, and the normalization factor is $N_{D}=-8\pi i$. The parameters $t=t'=0.5$ and $\left\{M_{1},M_{2},M_{3},M_{4}\right\}=\left\{-0.5,-0.2,0.2,0.5\right\}$ are used in the numerical simulation.

%{\cblue (1) My numerics does not work. If I cyclic the position operators then it seems like it works well, but if if I just use the expression as it is then it does not even get a real number. }

%{\cblue (2) My numerics for 3D class AIII must be wrong because the Chern marker is not 4-fold symmetric. There is some problem with the lattice model. }

\subsection{3D class DIII \label{sec:3D_class_DIII}}

A concrete example of 3D class DIII is the B-phase of superfluid $^{3}$He\cite{Balian63,Volovik09}. For the purpose of discussing the 2D class DIII case after a dimensional reduction, which will be addressed in Sec.~\ref{sec:2D_class_DIII}, we use the representation of $\Gamma$-matrices in the Bernevig-Hughes-Zhang (BHZ) model\cite{Bernevig06,Konig07} 
\begin{eqnarray}
\Gamma^{1\sim 5}=\left\{s_{x}\otimes\sigma_{z},s_{y}\otimes I,s_{z}\otimes I,s_{x}\otimes\sigma_{x},s_{x}\otimes\sigma_{y}\right\}.\;\;\;
\label{BHZ_Gamma_matrices}
\end{eqnarray}
The TR, PH, and chiral operators in this basis are $T=-iI\otimes\sigma_{y}K$, $C=s_{x}\otimes IK$, and $S=\Gamma^{5}$. Since we aim to demonstrate the topological marker on a lattice, we regularize the pairing terms of the B-phase of $^{3}$He on a square lattice by $\Delta k_{i}\rightarrow \Delta \sin k_{i}$, and likewisely the kinetic terms, and arrange our spinor according to Eq.~(\ref{BHZ_Gamma_matrices}) by $\eta_{\bf k}^{\dag}=(c_{\bf k\uparrow}^{\dag},c_{\bf -k\uparrow},c_{\bf k\downarrow}^{\dag},c_{\bf -k\downarrow})$. The leads to the parametrization of the Hamiltonian
\begin{eqnarray}
&&H=\sum_{\bf k}\eta_{\bf k}^{\dag}\left(\sum_{i=1}^{4}d_{i}\Gamma^{i}\right)\eta_{\bf k},\;\;\;
d_{1}=\Delta\sin k_{x},\;\;\;
d_{2}=\Delta\sin k_{y},
\nonumber \\
&&d_{3}=2t\left(\cos k_{x}+\cos k_{y}+\cos k_{z}\right)-\mu,\;\;\;
d_{4}=-\Delta\sin k_{z},
\label{3D_class_DIII_BHZ_parametrization}
\end{eqnarray}
so the unused Dirac matrix is the chiral operator $W=S=\Gamma^{5}$, and the normalization factor is $N_{D}=-8\pi i$. We then construct a square lattice model in a similar manner as Eq.~(\ref{regularization_k_to_site}), yielding 
\begin{eqnarray}
&&H=\sum_{i\sigma\delta}-t\left(c_{i\sigma}^{\dag}c_{i+\delta\sigma}+c_{i+\delta\sigma}^{\dag}c_{i\sigma}\right)-\mu \sum_{i\sigma}c_{i\sigma}^{\dag}c_{i\sigma}
\nonumber \\
&&+\sum_{i}\Delta\left(-ic_{i\uparrow}c_{i+x\uparrow}+ic_{i+x\uparrow}^{\dag}c_{i\uparrow}^{\dag}
+c_{i\uparrow}c_{i+y\uparrow}+c_{i+y\uparrow}^{\dag}c_{i\uparrow}^{\dag}\right)
\nonumber \\
&&+\sum_{i}\Delta\left(ic_{i\downarrow}c_{i+x\downarrow}-ic_{i+x\downarrow}^{\dag}c_{i\downarrow}^{\dag}
+c_{i\downarrow}c_{i+y\downarrow}+c_{i+y\downarrow}^{\dag}c_{i\downarrow}^{\dag}\right)
\nonumber \\
&&+\sum_{i}\Delta\left(ic_{i\uparrow}c_{i+z\downarrow}-ic_{i+z\downarrow}^{\dag}c_{i\uparrow}^{\dag}
+ic_{i\downarrow}c_{i+z\uparrow}-ic_{i+z\uparrow}^{\dag}c_{i\downarrow}^{\dag}\right).
\nonumber \\
\label{3D_class_DIII_lattice_model}
\end{eqnarray}
where $\delta=\left\{x,y,z\right\}$. We use $t=\Delta=0.5$ and $\left\{\mu_{1},\mu_{2},\mu_{3},\mu_{4}\right\}=\left\{4,3.5,2.5,2\right\}$ in the numerical calculation.

%{\cblue (1) My numerical calculation for 3D class DIII works well and is normalized correctly. }

\subsection{3D class AII}

The 3D class AII is relevant to prototype TIs such as Bi$_{2}$Se$_{3}$ and Bi$_{2}$Te$_{3}$. To draw relevance to real materials, we will use the model for the low energy sector described by the Dirac matrices\cite{Zhang09,Liu10}
\begin{eqnarray}
\Gamma_{1\sim 5}=\left\{\sigma^{x}\otimes\tau^{x},\sigma^{y}\otimes\tau^{x},\sigma^{z}\otimes\tau^{x},
I_{\sigma}\otimes\tau^{y},I_{\sigma}\otimes\tau^{z}\right\}.
\nonumber \\
\end{eqnarray}
The spinor is $\psi_{\bf k}=\left(c_{{\bf k}s\uparrow},c_{{\bf k}p\uparrow},c_{{\bf k}s\downarrow},c_{{\bf k}p\downarrow}\right)^{T}$, where $s$ and $p$ stand for the $P1_{-}^{+}$ and $P2_{+}^{-}$ orbitals in real materials. The low energy Hamiltonian given by the lowest order term in the ${\bf k\cdot p}$ theory 
\begin{eqnarray}
&&\hat{H}=\left(M+M_{1}k_{z}^{2}+M_{2}k_{x}^{2}+M_{2}k_{y}^{2}\right)\Gamma_{5}
+B_{0}\Gamma_{4}k_{z}
\nonumber \\
&&+A_{0}\left(\Gamma_{1}k_{y}-\Gamma_{2}k_{x}\right),
\label{3D_TI_H0_H1}
\end{eqnarray}
can be regularized on a cubic lattice, yielding\cite{Chen20_absence_edge_current}
\begin{eqnarray}
&&H=-\sum_{iI\sigma}\mu c_{iI\sigma}^{\dag}c_{iI\sigma}+\sum_{i\sigma}\tilde{M}\left\{c_{is\sigma}^{\dag}c_{is\sigma}-c_{ip\sigma}^{\dag}c_{ip\sigma}\right\}
\nonumber \\
&&+\sum_{iI}t_{\parallel}\left\{c_{iI\uparrow}^{\dag}c_{i+a\overline{I}\downarrow}
-c_{i+aI\uparrow}^{\dag}c_{i\overline{I}\downarrow}\right.
\nonumber \\
&&\left.-ic_{iI\uparrow}^{\dag}c_{i+b\overline{I}\downarrow}
+ic_{i+bI\uparrow}^{\dag}c_{i\overline{I}\downarrow}+h.c.\right\}
\nonumber \\
&&+\sum_{i\sigma}t_{\perp}\left\{-c_{is\sigma}^{\dag}c_{i+cp\sigma}+c_{i+cs\sigma}^{\dag}c_{ip\sigma}+h.c.\right\}
\nonumber \\
&&-\sum_{i\sigma}M_{1}\left\{c_{is\sigma}^{\dag}c_{i+cs\sigma}-c_{ip\sigma}^{\dag}c_{i+cp\sigma}+h.c.\right\}
\nonumber \\
&&-\sum_{i\delta\sigma}M_{2}\left\{c_{is\sigma}^{\dag}c_{i+\delta s\sigma}-c_{ip\sigma}^{\dag}c_{i+\delta p\sigma}+h.c.\right\},
\label{3DTIFMM_Hamiltonian}
\end{eqnarray}
where $\tilde{M}=M+2M_{1}+4M_{2}$, $t_{\parallel}=A_{0}/2$, $t_{\perp}=B_{0}/2$, $I=\left\{s,p\right\}$ and $\overline{I}=\left\{p,s\right\}$ are the orbital indices, $\delta=\left\{a,b,c\right\}$ denotes the lattice constants, and $\sigma=\left\{\uparrow,\downarrow\right\}$ is the spin index. From Eq.~(\ref{3D_TI_H0_H1}), it is clear that the Dirac matrix that has not been used is $W=\Gamma_{3}$, and the normalization factor is $N_{D}=-8\pi i$. In the numerical calculation, we use the parameters $t_{\parallel}=t_{\perp}=M_{1}=M_{2}=1$, and four values $\left\{-2,-1,1,2\right\}$ for the mass term $M$ to capture the critical behavior near $M_{c}=0$. 

%{\cblue (1) My numerical result for 3D class AII works well, with the correct normalization factor. }

\subsection{3D class CII}

The minimal model of 3D class CII is a $8\times 8$ Dirac model\cite{Schnyder08}, where the seven $\Gamma$-matrices are given by\cite{Ryu10} 
\begin{eqnarray}
&&\Gamma^{a}=\Gamma_{4\times 4}^{a}\otimes\eta_{x}\;,\;\;\;{\rm for}\;a=1\sim 4\;,
\nonumber \\
&&\Gamma^{5}=I_{4\times 4}\otimes\eta_{y}\;,\;\;\;\Gamma^{6}=I_{4\times 4}\otimes\eta_{z}\;,
\nonumber \\
&&\Gamma^{7}=(-i)^{3}\Gamma^{1}\Gamma^{2}...\Gamma^{6}\;,
\end{eqnarray}
where $\Gamma_{4\times 4}^{a}$ are those in Eq.~(\ref{3D_class_AIII_gamma_matrices}). The chiral symmetry is implemented by  $\Gamma^{6}=S$. The Hamiltonian expressed in terms of the other six $\Gamma$-matrices has a block-off-diagonal form
\begin{eqnarray}
&&H({\bf k})=\sum_{i=1,2,3,4,5,7}d_{i}\Gamma^{i}=\left(
\begin{array}{cccc}
 & & D_{11} & D_{12} \\
 & & D_{21} & D_{22} \\
D_{11}^{\ast} & D_{12} & & \\
D_{21} & D_{22}^{\ast} & & 
\end{array}
\right)\;,
\nonumber \\
&&D_{11}=\left(
\begin{array}{cc}
h & \\
 & h 
\end{array}\right)=-D_{22}^{\ast}\;,\;\;\;
D_{12}=\left(
\begin{array}{cc}
g & f \\
f^{\ast} & -g^{\ast} 
\end{array}\right)=D_{21}^{\dag}.
\nonumber \\
\end{eqnarray}
Within linear Dirac model, we consider
\begin{eqnarray}
&&f=d_{1}-id_{2}=Ak_{x}-iAk_{y},\;\;\;g=d_{3}+id_{7}=d_{3}=Ak_{z},
\nonumber \\
&&h=d_{4}-id_{5}=d_{4}=M.
\end{eqnarray}
The Dirac matrices that are omitted in the Hamiltonian are $\left\{\Gamma^{5},\Gamma^{6},\Gamma^{7}\right\}$, so we choose $W=\Gamma^{5}\Gamma^{6}\Gamma^{7}$, and the normalization factor is $N_{D}=-4\pi i/c$. Since this symmetry class is less explored in the literature, and given that complexity involved in this $8\times 8$ model, the examination of the lattice model is left for future investigations.

%We then procees to regularize the Hamiltonian on the whole BZ and construct a lattice model according to Eqs.~(\ref{k_to_sink}) and (\ref{regularization_k_to_site}) by simply labeling the eight degrees of freedom on site $i$ by $\left\{c_{i1},c_{i2}... c_{i8}\right\}$, yielding 
%\begin{eqnarray}
%&&H=\sum_{i}(M+6t')\left\{c_{i1}^{\dag}c_{i5}+c_{i2}^{\dag}c_{i6}-c_{i3}^{\dag}c_{i7}-c_{i4}^{\dag}c_{i8}\right\}
%\nonumber \\
%&&-t'\sum_{i\delta}\left\{c_{i1}^{\dag}c_{i+\delta 5}+c_{i+\delta 1}^{\dag}c_{i 5}+c_{i2}^{\dag}c_{i+\delta 6}+c_{i+\delta 2}^{\dag}c_{i 6}\right.
%\nonumber \\
%&&\left.-c_{i3}^{\dag}c_{i+\delta 7}-c_{i+\delta 3}^{\dag}c_{i 7}-c_{i4}^{\dag}c_{i+\delta 8}-c_{i+\delta 4}^{\dag}c_{i8}\right.
%\nonumber \\
%&&+t\sum_{i}\left\{c_{i1}^{\dag}c_{i+x8}+c_{i+x1}^{\dag}c_{i 8}+c_{i2}^{\dag}c_{i+x7}+c_{i+x2}^{\dag}c_{i7}\right.
%\nonumber \\
%&&+c_{i3}^{\dag}c_{i+x6}+c_{i+x3}^{\dag}c_{i6}+c_{i4}^{\dag}c_{i+x5}+c_{i+x4}^{\dag}c_{i5}
%\nonumber \\
%&&-c_{i1}^{\dag}c_{i+y8}+c_{i+y1}^{\dag}c_{i8}+c_{i2}^{\dag}c_{i+y7}-c_{i+y2}^{\dag}c_{i7}
%\nonumber \\
%&&-c_{i3}^{\dag}c_{i+y6}+c_{i+y3}^{\dag}c_{i6}+c_{i4}^{\dag}c_{i+y5}-c_{i+y4}^{\dag}c_{i5}
%\nonumber \\
%&&+c_{i1}^{\dag}c_{i+z7}+c_{i+z1}^{\dag}c_{i7}-c_{i2}^{\dag}c_{i+z8}-c_{i+z2}^{\dag}c_{i8}
%\nonumber \\
%&&\left.+c_{i3}^{\dag}c_{i+z5}+c_{i+z3}^{\dag}c_{i5}-c_{i4}^{\dag}c_{i+z6}-c_{i+z4}^{\dag}c_{i6}\right\}+h.c.
%\end{eqnarray}

%{\cblue (1) I did the numerics accordingly and it does not work. not sure why. Need to check my program topologicalmarker-3D-CII.nb }

\subsection{3D class CI}

For 3D class CI, we resort to the $8\times 8$ Hamiltonian expanded by 4 out of the 7 Dirac matrices\cite{Schnyder08} 
\begin{eqnarray}
&&H=\sum_{i=1}^{4}d_{i}\Gamma_{i}=\left(
\begin{array}{cc}
 & D \\
D^{\dag} & 
\end{array}
\right),
\nonumber \\
&&D=\left(
\begin{array}{cc}
 & D_{12} \\
D_{21} & 
\end{array}
\right)
=\left(
\begin{array}{cccc}
 & & f^{\ast} & -g \\
 & & -g^{\ast} & -f \\
-f^{\ast} & g & & \\
g^{\ast} & f & &
\end{array}
\right)\;,
\label{3D_class_CI_D}
\end{eqnarray}
with $f=d_{1}-id_{2}$ and $g=d_{3}+id_{4}$. The $8\times 8$ TR and PH operators are $T=I\otimes I\otimes \sigma_{x}K$ and $C=I\otimes I\otimes(-i\sigma_{y})K$, which require $d_{1}\sim d_{3}$ to be odd in momentum, and $d_{4}$ to be the mass term that is even in momentum. Note that we do not need to know the explicit form of the unused $\Gamma_{5}\sim\Gamma_{7}$ matrices to calculate their product $W=\Gamma_{5}\Gamma_{6}\Gamma_{7}$, since we know that 
\begin{eqnarray}
\Gamma_{1}\Gamma_{2}\Gamma_{3}\Gamma_{4}\Gamma_{5}\Gamma_{6}\Gamma_{7}=
\Gamma_{1}\Gamma_{2}\Gamma_{3}\Gamma_{4}W=c\,I_{8\times 8}.
\end{eqnarray}
Because $\Gamma_{1}\Gamma_{2}\Gamma_{3}\Gamma_{4}={\rm diag}(-1,1)_{8\times 8}$, one sees that $W={\rm diag}(-c,c)_{8\times 8}=-cS$ is given by the chiral operator, and $N_{D}=-4\pi i/c$. The exploration of the lattice model corresponding to this $8\times 8$ Hamiltonian will be left for further investigations.

%Secondly, the form of the PH operator $C$ indicates that the first 4 elements of the spinor is the particle channel and the last 4 elements are the hole channel. In addition, the mass term $d_{4}\Gamma_{4}$ in Eq.~(\ref{3D_class_CI_D}) implies that the elements $(1,8)$, $(2,7)$, $(3,6)$, and $(4,5)$ in the spinor are pairs of the same flavor, hence we arrange the spinor by
%\begin{eqnarray}
%\psi^{\dag}=(c_{1\bf k}^{\dag},c_{2\bf k}^{\dag},c_{3\bf k}^{\dag},c_{4\bf k}^{\dag},c_{4\bf -k},c_{3\bf -k},c_{2\bf -k},c_{1\bf -k}).
%\end{eqnarray}
%where the subscript $1\sim 4$ are the flavor index that may be spin, orbit, etc, whose precise interpretation does not influence our conclusions.

%{\cblue (1) As of Aug 19, 2022, I have problems to assign the spinor of 3D class CI. It seems like the fact that the mass term is even but the pairing terms is odd yet they enter the same matrix element has some problems. }

\section{Topological markers in two dimensions}

\begin{figure}[ht]
\begin{center}
\includegraphics[clip=true,width=0.99\columnwidth]{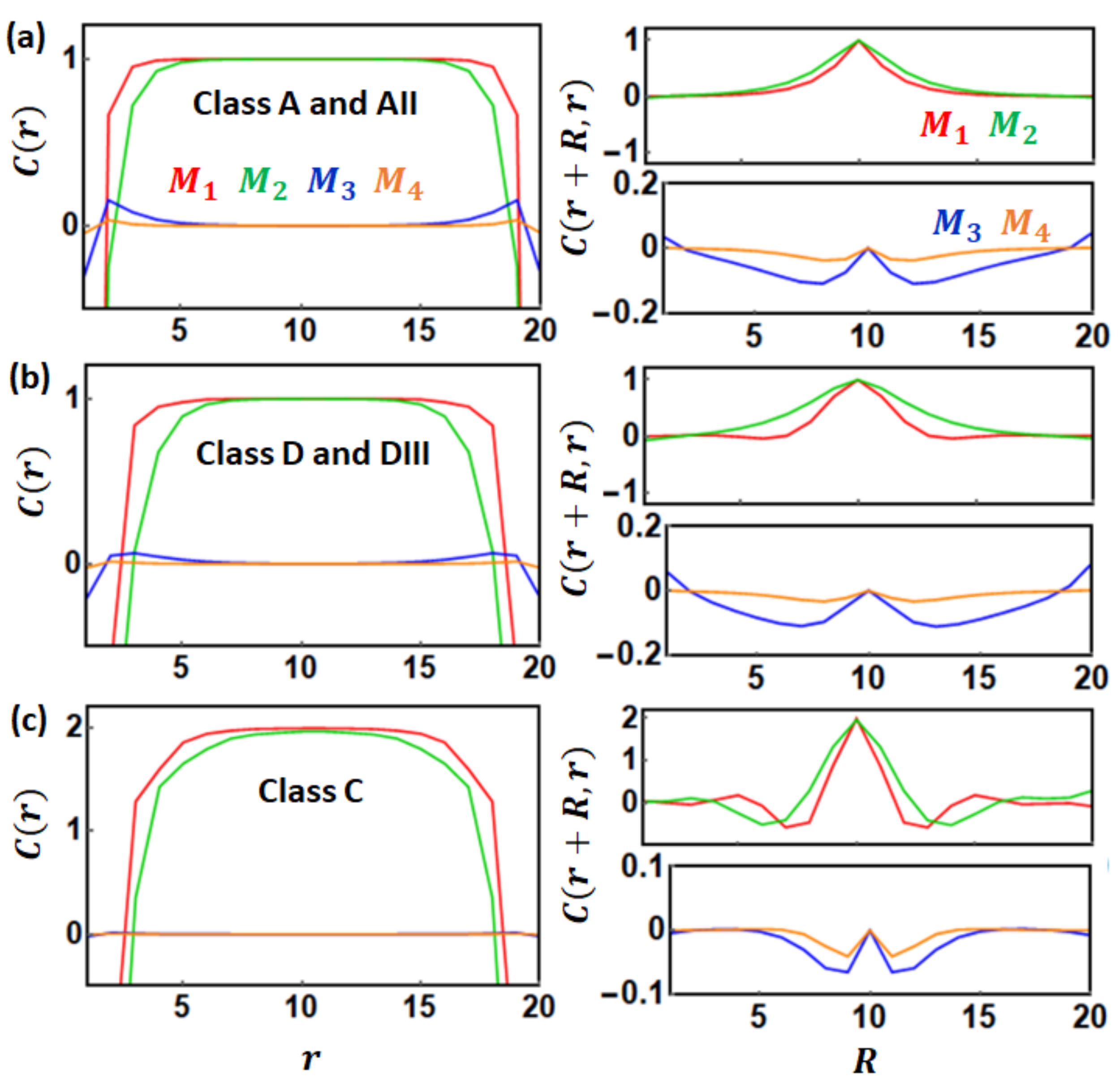}
\caption{Local (left column) and nonlocal (right column) topological markers for prototype square lattice models in the five topologically nontrivial symmetry classes in 2D, including (a) the Chern insulator in class A and the BHZ model in class AII that have identical results, (b) the chiral $p$-wave SC in class D and helical $p$-wave SC in class DIII that have identical results, (c) and a quadratic band crossing model in class C. } 
\label{fig:2D_results}
\end{center}
\end{figure}

For 2D TIs and TSCs, the topological operator reads
\begin{eqnarray}
{\hat{\cal C}}_{2D}=N_{D}W\left[Q{\hat x}P{\hat y}Q-P{\hat x}Q{\hat y}P\right].
\end{eqnarray}
We find that there are only two kinds of topological markers in 2D: For classes A, C, and D that break TR symmetry, the topological marker is the Chern marker\cite{Bianco11} described by $W\propto I$. On the other hand, the TR-symmetric classes AII and DIII are described by the spin Chern number, yielding a spin Chern marker $W\propto\sigma_{z}$ that counts the difference between the spin up and down channels. The numerical calculation using $20\times 20$ lattices for all the 5 nontrivial symmetry classes is presented in Fig.~\ref{fig:2D_results}, as detailed below.

\subsection{2D class A \label{sec:2D_class_A}}

The minimal model of 2D class A is expanded by all three components of Pauli matrices $H=\sum_{i=1}^{3}d_{i}\sigma_{i}$, and the model regularized on the whole BZ
\begin{eqnarray}
&&d_{1}=A\sin k_{x},\;\;\;d_{2}=A\sin k_{y},
\nonumber \\
&&d_{3}=M+4B-2B\cos k_{x}-2B\cos k_{y},
\end{eqnarray}
gives the Chern insulator, whose lattice model has been given previously\cite{Chen20_absence_edge_current,Molignini22_Chern_marker}. Since all Pauli matrices are used, we have $W=I$, and the normalization factor is $N_{D}=2\pi i$. The projection to lattice sites $|{\bf r}\rangle$ is equivalent to the original Chern marker that have been intensively studied\cite{Bianco11}, and the off-diagonal elements has been called nonlocal Chern marker\cite{Molignini22_Chern_marker}. Nevertheless, for sake of completeness of the presentation, we perform simulations using the parameters $t=A/2=1$, $t'=B=1$, $\left\{M_{1},M_{2},M_{3},M_{4}\right\}=\left\{-2,-0.8,0.8,2\right\}$.

\subsection{2D class D \label{sec:2D_class_D}}

A concrete system that realizes 2D class D is the spinless chiral $p$-wave SC\cite{Schnyder08}, described by the lattice Hamiltonian 
\begin{eqnarray}
&&H=\sum_{i\delta}t\left(c_{i}^{\dag}c_{i+\delta}+c_{i+\delta}^{\dag}c_{i}\right)-\mu \sum_{i}c_{i}^{\dag}c_{i}
\nonumber \\
&&+\sum_{i}\Delta \left(-ic_{i}c_{i+x}+ic_{i+x}^{\dag}c_{i}^{\dag}+c_{i}c_{i+y}+c_{i+y}^{\dag}c_{i}^{\dag}\right),
\label{chiral_pwave_lattice_model}
\end{eqnarray}
where $\delta=\left\{x,y\right\}$, and $c_{i}$ is the spinless fermion operator at site $i$. All three components of Pauli matrices are used, so we also use $W=I$ and $N_{D}=2\pi i$. The parameters examined are $t=-1$, $\Delta=0.5$, $\left\{\mu_{1},\mu_{2},\mu_{3},\mu_{4}\right\}=\left\{-3,-3.7,-4.3,-5\right\}$.

\subsection{2D class C}

The minimal model of 2D class C is a $2\times 2$ Dirac model\cite{Chen19_universality}, where the off-diagonal pairing term around the HSP can be expanded by $d_{1}-id_{2}=k_{+}^{n_{+}}k_{-}^{n_{-}}$, with $k_{\pm}=k_{x}\pm ik_{y}$. We consider a spinless model of particle-hole basis $\eta_{\bf k}^{\dag}=(c_{\bf k}^{\dag},c_{\bf -k})$, where the PH operator $C=\sigma_{y}K$ requires all $d_{i}({\bf k})$ to be even in momentum, i.e., the model has even-order band crossing at TPTs. For concreteness, we choose to examine the model with the power $n_{+}=0$, $n_{-}=2$, which may be regularized on a lattice to give the following pairing term 
\begin{eqnarray}
\Delta_{\bf k}=\Delta(2\cos k_{x}-2\cos k_{y}-2i\sin k_{x}\sin k_{y}).
\end{eqnarray}
This leads us to consider the lattice model that contains both nearest- and next-nearest-neighbor pairings of the same amplitude but a phase difference
\begin{eqnarray}
&&H=\sum_{i\delta}t\left(c_{i}^{\dag}c_{i+\delta}+c_{i+\delta}^{\dag}c_{i}\right)-\mu \sum_{i}c_{i}^{\dag}c_{i}
\nonumber \\
&&+\Delta\sum_{i,\sigma=\pm}\left(c_{i}c_{i+\sigma x}+c_{i+\sigma x}^{\dag}c_{i}^{\dag}-c_{i}c_{i+\sigma y}-c_{i+\sigma y}^{\dag}c_{i}^{\dag}\right)
\nonumber \\
&&+\frac{\Delta}{2}\sum_{i}\left\{-ic_{i}c_{i+x+y}+ic_{i}c_{i+x-y}+ic_{i}c_{i-x+y}-ic_{i}c_{i-x-y}\right.
\nonumber \\
&&\left.+ic_{i+x+y}^{\dag}c_{i}^{\dag}-ic_{i+x-y}^{\dag}c_{i}^{\dag}
-ic_{i-x+y}^{\dag}c_{i}^{\dag}+ic_{i-x-y}^{\dag}c_{i}^{\dag}\right\}.
\label{2DclassC_lattice_model}
\end{eqnarray}
Since the model already uses all the Dirac matrices, one has $W=I$ and $N_{D}=2\pi i$. We use the parameters $t=1$, $\Delta=0.5$, and $\left\{\mu_{1},\mu_{2},\mu_{3},\mu_{4}\right\}=\left\{2,3,4.3,5\right\}$. Note that deep inside the bulk and in the topologically nontrivial phase, we obtain ${\cal C}({\bf r})\approx 2$, consistent with that expected from the quadratic band crossing\cite{Chen19_universality}.

\subsection{2D class AII \label{sec:2D_class_AII}}

For 2D class AII that has TR symmetry, we consider the prototype BHZ model with spinful $s$ and $p$ orbitals $\psi=\left(s\uparrow,p\uparrow,s\downarrow,p\downarrow\right)^{T}$, which uses the Dirac matrices of this model already given in Eq.~(\ref{BHZ_Gamma_matrices}). Interestingly, the unused Dirac matrices combined to give the spin operator $W=\Gamma_{3}\Gamma_{4}=iI\otimes\sigma^{z}$, and the normalization factor is found to be $N_{D}=\pi$. The diagonal and off-diagonal elements of this topological operator has been called local and nonlocal spin Chern markers previously, whose validity applied to the BHZ model has been elaborated explicitly. The numerical result is exactly the same as the Chern insulator in Sec.~\ref{sec:2D_class_A} if the same parameters are used, as expected since BHZ model is equivalently two copies of Chern insulators, one for each spin species.

\subsection{2D class DIII \label{sec:2D_class_DIII}}

The lattice model of 2D class DIII can be obtained from that of the 3D class DIII presented in Sec.~\ref{sec:3D_class_DIII} via a dimensional reduction, which turns off all the $\sin k_{z}$ and $\cos k_{z}$ terms in the momentum space Hamiltonian in Eq.~(\ref{3D_class_DIII_BHZ_parametrization}), and equivalently all the terms that contain $c_{i+z\sigma}$ or $c_{i+z\sigma}^{\dag}$ in the lattice Hamiltonian in Eq.~(\ref{3D_class_DIII_lattice_model}). Since we arrange our spinor by $\eta_{\bf k}^{\dag}=(c_{\bf k\uparrow}^{\dag},c_{\bf -k\uparrow},c_{\bf k\downarrow}^{\dag},c_{\bf -k\downarrow})$, this results in a Hamiltonian that is block-diagonal, where each block corresponds to a chiral $p$-wave SC addressed in Sec.~\ref{sec:2D_class_D} for one spin species, and therefore describes a helical $p$-wave SC. 

%This can be easily seen by recognizing that the triplet pairing channel of spin up in Eq.~(\ref{3D_class_DIII_lattice_model}) is the same as the pairing channel of chiral $p$-wave SC in Eq.~(\ref{chiral_pwave_lattice_model}), and so is the triplet pairing channel of spin down in Eq.~(\ref{3D_class_DIII_lattice_model}) except the $x$-direction changes sign. 

The 3D class DIII model in Sec.~\ref{sec:3D_class_DIII} already omits the $\Gamma^{5}$ component, and the dimensional reduction to 2D turns off the $\Gamma^{4}$ matrix, so the unused $\Gamma$-matrices multiplied together $W=\Gamma^{3}\Gamma^{5}=iI\otimes\sigma^{z}$ gives the spin polarization operator, indicating that the topological operator is precisely the spin Chern operator discussed in Sec.~\ref{sec:2D_class_AII} with a normalization factor $N_{D}=\pi$. Physically, this means that the topological invariant is given by the difference between the Chern number of the spin up chiral $p$-wave SC and that of the spin down component. As a result, the spin Chern marker of the helical $p$-wave SC is identical to the Chern marker of the chiral $p$-wave SC given in Sec.~\ref{sec:2D_class_D} at the same parameters. 

%since the two spin degrees of freedom are separated due to the Block-diagonal Hamiltonian yet together they preserve TR symmetry. 

\section{Topological markers in one dimension}

\begin{figure}[ht]
\begin{center}
\includegraphics[clip=true,width=0.99\columnwidth]{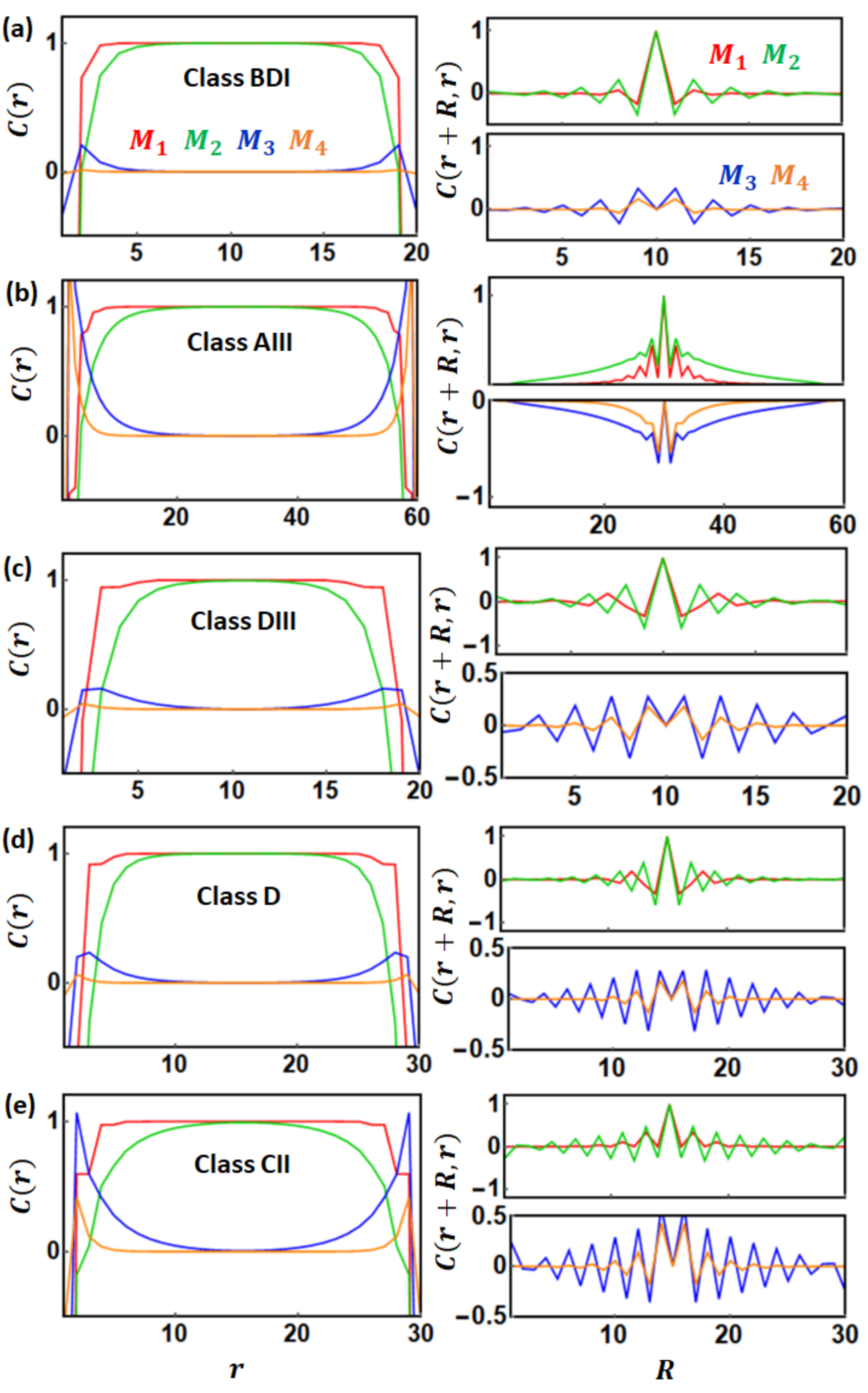}
\caption{Local (left column) and nonlocal (right column) topological markers for the lattice models in the five topologically nontrivial symmetry classes in 1D, including (a) the SSH model in class BDI, (b) a regularized lattice model in class AIII, (c) a class DIII model obtained from dimensional reduction, (d) the Kitaev $p$-wave SC chain in class D, and (e) a regularized lattice model in class CII. } 
\label{fig:1D_results}
\end{center}
\end{figure}

The topological operator in 1D takes the form
\begin{eqnarray}
{\hat{\cal C}}_{1D}=N_{D}W\left[Q{\hat x}P+P{\hat x}Q\right].
\end{eqnarray}
Interestingly, for all the symmetry classes in 1D that preserve chiral symmetry (AIII, BDI, CII, DIII), the product of unused Dirac matrices is always proportional to the chiral operator $W\propto S$, whereas the class D that does not preserve chiral symmetry has a different interpretation of $W$. The numerical results for these 5 classes are given in Fig.~\ref{fig:1D_results} and are described in detail below.

\subsection{1D class BDI}

For 1D class BDI, we use the prototype spinless Su-Schrieffer-Heeger (SSH) model as an example, which is described by the lattice Hamiltonian~\cite{Su79} 
\begin{eqnarray}
{\cal H}_{0}&=&\sum_{i}(t+\delta t)c_{Ai}^{\dag}c_{Bi}+(t-\delta t)c_{Ai+1}^{\dag}c_{Bi}+h.c.
\end{eqnarray}
where $c_{Ai}$ and $c_{Bi}$ are the fermion annihilation operators on sublattice $A$ and $B$ in the unit cell $i$, respectively, and $t\pm\delta t$ are the alternating hopping amplitudes. The $2\times 2$ Hamiltonian expressed in momentum space with the basis $(c_{Ak},c_{Bk})$ is expanded by $\sigma_{x}$ and $\sigma_{y}$, so the only Pauli matrix that has not been used is the chiral operator $W=S=\sigma_{z}$, and the normalization factor is unity $N_{D}=1$. Note that as discussed after Eq.~(\ref{mdin_psiipsi}), to realize the position operator ${\hat x}$ as a diagonal matrix, the $A$ and $B$ sublattices within the same unit cell located at $i$ are assigned with the same position $x_{i}$, even though they are frequently drawn as a certain distance apart. 
We choose parameters $t=1$, $\left\{\delta t_{1},\delta t_{2},\delta t_{3},\delta t_{4}\right\}=\left\{-0.5,-0.2,0.2,0.5\right\}$ and 20 lattice sites in the numerical simulation.

\subsection{1D class AIII}

The low energy linear Dirac model\cite{Chen19_universality} for 1D class AIII can be expanded by $H({\bf k})=Ak_{x}\sigma_{x}+M\sigma_{z}$, with the unused Pauli matrix being the chiral symmetry operator $W=S=\sigma_{y}$ and the normalization factor $N_{D}=1$. The regularization in Eqs.~(\ref{k_to_sink}) and (\ref{regularization_k_to_site}) leads to a lattice model
\begin{eqnarray}
&&H=\sum_{i}(M-2t')\left(c_{i1}^{\dag}c_{i1}-c_{i2}^{\dag}c_{i2}\right)
\nonumber \\
&&+\sum_{i}t'\left(c_{i1}^{\dag}c_{i+x1}+c_{i+x1}^{\dag}c_{i1}+c_{i2}^{\dag}c_{i+x2}+c_{i+x2}^{\dag}c_{i2}\right)
\nonumber \\
&&+\sum_{i}t\left(-ic_{i1}^{\dag}c_{i+x2}+ic_{i+x2}^{\dag}c_{i1}-ic_{i2}^{\dag}c_{i+x1}+ic_{i+x1}^{\dag}c_{i2}\right).
\nonumber \\
\end{eqnarray}
We have used $t=1$, $t'=0.4$, $\left\{M_{1},M_{2},M_{3},M_{4}\right\}=\left\{0.6,0.2,-0.2,-0.6\right\}$ and 60 lattice sites in the numerical simulation.

\subsection{1D class DIII}

We construct a lattice model of 1D class DIII by performing dimensional reduction twice on the 3D class DIII model in Sec.~\ref{sec:3D_class_DIII}, which is done by turning off all the $\left\{\sin k_{y},\sin k_{z},\cos k_{y},\cos k_{z}\right\}$ terms in Eq.~(\ref{3D_class_DIII_BHZ_parametrization}), and analogously turning off all the $\left\{c_{i+y\sigma},c_{i+y\sigma}^{\dag},c_{i+z\sigma},c_{i+z\sigma}^{\dag}\right\}$ terms in Eq.~(\ref{3D_class_DIII_lattice_model}). The resulting Hamiltonian omits $\left\{\Gamma^{2},\Gamma^{4},\Gamma^{5}\right\}$ matrices defined in Eq.~(\ref{BHZ_Gamma_matrices}), so $W=\Gamma^{2}\Gamma^{4}\Gamma^{5}=S$ is given by the chiral operator, and the normalization factor is $N_{D}=i/2$. We use the parameters $t=1$, $\Delta=0.5$, $\left\{\mu_{1},\mu_{2},\mu_{3},\mu_{4}\right\}=\left\{1,1.8,2.2,3\right\}$ and 20 lattice sites in the numerical calculation.

\subsection{1D class D}

For 1D class D, we examine the spinless Kitaev $p$-wave SC chain described by\cite{Kitaev01}
\begin{eqnarray}
&&H=\sum_{i}t\left(c_{i}^{\dag}c_{i+1}+c_{i+1}^{\dag}c_{i}\right)-\mu \sum_{i}c_{i}^{\dag}c_{i}
\nonumber \\
&&+\sum_{i}\Delta \left(c_{i}c_{i+1}+c_{i+1}^{\dag}c_{i}^{\dag}\right),
\label{Majorana_lattice_model}
\end{eqnarray}
where $c_{i}$ is the spinless fermion annihilation operator at site $i$. The Hamiltonian in momentum space in the basis of $(c_{k},c_{-k}^{\dag})^{T}$ is spanned by $\sigma_{z}$ and $\sigma_{y}$, so the only Pauli matrix that has not been used is $W=\sigma_{x}$, and $N_{D}=1$. A 30-site lattice with the parameters $t=1$, $\Delta=0.5$, $\left\{\mu_{1},\mu_{2},\mu_{3},\mu_{4}\right\}=\left\{1,1.8,2.2,3\right\}$ is used in the numerical simulation.

%{\cblue (Well, this is strange, because if $WH(k)W^{-1}=-H(k)$ is satisfied by $W=\sigma_{x}=S$ then it means Kitaev chain should have chiral symmetry, but then it wouldn't be class D anymore? I think this is because $\sigma_{x}=KC$ but $K$ cannot be interpreted as TR symmetry because the spinor is a particle-hole basis, hence $\sigma_{x}=KC\neq TC=S$ should not be interpreted as a chiral operator.)}

\subsection{1D class CII}

For 1D class CII, we adopt the $\Gamma$-matices\cite{Zhao14}
\begin{eqnarray}
\Gamma^{a}=\left\{\sigma_{x}\otimes\tau_{z},\sigma_{y}\otimes\tau_{z},
I\otimes\sigma_{x},I\otimes\tau_{y},\sigma_{z}\otimes\tau_{z}\right\}\;,
\end{eqnarray}
and the TR and PH operators are interpreted by $T=\sigma_{y}\otimes IK$ and $C=I\otimes\tau_{y}K$. The minimal model in momentum space is $H=d_{2}\Gamma^{2}+d_{3}\Gamma^{3}=Ak\Gamma^{2}+M\Gamma^{3}$, and we denote the spinor by $\psi_{k}^{\dag}=(c_{1k}^{\dag},c_{2k}^{\dag},c_{3k}^{\dag},c_{4k}^{\dag})$ where the four degrees of freedom are enumerated by $1\sim 4$. The regularization on a lattice gives
\begin{eqnarray}
&&H=\sum_{i}t\left\{-c_{i1}^{\dag}c_{i+x2}+c_{i2}^{\dag}c_{i+x1}+c_{i3}^{\dag}c_{i+x4}-c_{i4}^{\dag}c_{i+x3}\right\}
\nonumber \\
&&+\sum_{i}(-t')\left\{c_{i1}^{\dag}c_{i+x3}+c_{i3}^{\dag}c_{i+x1}+c_{i2}^{\dag}c_{i+x4}+c_{i4}^{\dag}c_{i+x2}\right\}.
\nonumber \\
&&+\sum_{i}(M+2t')\left\{c_{i1}^{\dag}c_{i3}+c_{i2}^{\dag}c_{i4}\right\}+h.c.
\end{eqnarray}
The unused Dirac matrices multiplied together is proportional to the chiral operator $W=\Gamma^{1}\Gamma^{4}\Gamma^{5}=-iS$, and the normalization factor is $N_{D}=i/2$. We use the parameters $t=1$, $t'=0.5$, $\left\{M_{1},M_{2},M_{3},M_{4}\right\}=\left\{-1,-1.8,-2.2,-3\right\}$ and a 30-site lattice in the simulation.

\section{Conclusions}

In summary, we show that topological marker can be constructed in a unified manner for TIs and TSCs in any dimension and symmetry class. The central object in our formalism is the topological operator in Eq.~(\ref{topological_operator}) derived from the universal topological invariant in momentum space, which takes the form of alternating projectors to the lattice eigenstates and the position operators, multiplied by the Dirac matrices that are omitted in the Hamiltonian. The ${\bf r}$-th diagonal element of the topological operator gives the local topological marker that recovers the topological invariant for lattice sites deep inside the bulk. In addition, the $({\bf r+R,r})$-th off-diagonal element yields a nonlocal topological marker that decays with ${\bf R}$, whose decay length diverges at TPT and may be interpreted as a Wannier state correlation function, thereby serving as a faithful correlator to identify TPTs in real space. The topological operator is constructed explicitly for each of the 15 topologically nontrivial symmetry classes in 1D to 3D. For 13 out of these 15 cases, we perform numerical calculation on concrete lattice models to demonstrate the validity of our topological marker, which cover a great number of prototype TIs and TSCs including the SSH model, Majorana chain, Chern insulator, BHZ model, chiral and helical $p$-wave SCs, 3D TR invariant TIs, lattice model of $^{3}$He B-phase, among many others, suggesting the ubiquity of our formalism.

Our results point to many open questions that remain to be clarified. Firstly, it is known that the deviation of the Chern marker at the boundary sites of the 2D lattice, which occur even if periodic boundary condition is imposed, comes from the fact that the position operators $\hat{i}$ in Eq.~(\ref{topological_operator}) do not respect the translational invariance, and may be cured by exponentiating the position operators\cite{Prodan10,Prodan10_2,Prodan11}. Whether such an exponentiating trick can be generally applied to topological markers in any dimension and symmetry class, and whether it has some non-commutative interpretation on the topological order in general, remain to be investigated. Secondly, a major category of topological materials that we did not address are the topological semimetals in $D$-dimension, such as graphene in 2D, in which the momentum space topological invariant is the wrapping number in Eq.~(\ref{wrapping_number_formula}) but integrated over a $(D-1)$-dimensional surface enclosing a nodal point. Because the momentum integration is one dimension lower, it is unclear to us at present whether the projector algebra in Sec.~\ref{sec:topological_operators} still applies, or how it may be modified to construct a topological marker for semimetals. Thirdly, concerning the experimental measurement of the topological marker, it has been pointed out that the Chern marker in 2D TR-breaking systems can be measured by circular dichroism\cite{Molignini22_Chern_marker}, and the spin-Chern marker in 2D TR-symmetric systems can be detected by spin-resolved circular dichroism\cite{Chen22_spin_Chern_marker}, both are due to the linear response of valence electrons to polarized electric field. However, because TIs and TSCs in other dimensions do not respond linearly to the electric field\cite{Bernevig13}, it remains to be investigated whether higher order responses can help to extract the topological marker in other dimensions, or if one has to resort to some other kinds of experimental protocol. Finally, an obvious question is whether our universal topological marker can still describe systems that are beyond the paradigm of Dirac models, such as 2D class AII systems with spin-orbit coupling\cite{Kane05_2}. All these open questions, together with the applications of the universal topological marker on issues such as real space inhomogeneity and topological quantum criticality, are intriguing subjects that await to be explored.

\bibliography{Literatur}

\begin{thebibliography}{46}%
\makeatletter
\providecommand \@ifxundefined [1]{%
 \@ifx{#1\undefined}
}%
\providecommand \@ifnum [1]{%
 \ifnum #1\expandafter \@firstoftwo
 \else \expandafter \@secondoftwo
 \fi
}%
\providecommand \@ifx [1]{%
 \ifx #1\expandafter \@firstoftwo
 \else \expandafter \@secondoftwo
 \fi
}%
\providecommand \natexlab [1]{#1}%
\providecommand \enquote  [1]{``#1''}%
\providecommand \bibnamefont  [1]{#1}%
\providecommand \bibfnamefont [1]{#1}%
\providecommand \citenamefont [1]{#1}%
\providecommand \href@noop [0]{\@secondoftwo}%
\providecommand \href [0]{\begingroup \@sanitize@url \@href}%
\providecommand \@href[1]{\@@startlink{#1}\@@href}%
\providecommand \@@href[1]{\endgroup#1\@@endlink}%
\providecommand \@sanitize@url [0]{\catcode `\\12\catcode `\$12\catcode
  `\&12\catcode `\#12\catcode `\^12\catcode `\_12\catcode `\%12\relax}%
\providecommand \@@startlink[1]{}%
\providecommand \@@endlink[0]{}%
\providecommand \url  [0]{\begingroup\@sanitize@url \@url }%
\providecommand \@url [1]{\endgroup\@href {#1}{\urlprefix }}%
\providecommand \urlprefix  [0]{URL }%
\providecommand \Eprint [0]{\href }%
\providecommand \doibase [0]{http://dx.doi.org/}%
\providecommand \selectlanguage [0]{\@gobble}%
\providecommand \bibinfo  [0]{\@secondoftwo}%
\providecommand \bibfield  [0]{\@secondoftwo}%
\providecommand \translation [1]{[#1]}%
\providecommand \BibitemOpen [0]{}%
\providecommand \bibitemStop [0]{}%
\providecommand \bibitemNoStop [0]{.\EOS\space}%
\providecommand \EOS [0]{\spacefactor3000\relax}%
\providecommand \BibitemShut  [1]{\csname bibitem#1\endcsname}%
\let\auto@bib@innerbib\@empty
%</preamble>
\bibitem [{\citenamefont {Hasan}\ and\ \citenamefont {Kane}(2010)}]{Hasan10}%
  \BibitemOpen
  \bibfield  {author} {\bibinfo {author} {\bibfnamefont {M.~Z.}\ \bibnamefont
  {Hasan}}\ and\ \bibinfo {author} {\bibfnamefont {C.~L.}\ \bibnamefont
  {Kane}},\ }\href {\doibase 10.1103/RevModPhys.82.3045} {\bibfield  {journal}
  {\bibinfo  {journal} {Rev. Mod. Phys.}\ }\textbf {\bibinfo {volume} {82}},\
  \bibinfo {pages} {3045} (\bibinfo {year} {2010})}\BibitemShut {NoStop}%
\bibitem [{\citenamefont {Qi}\ and\ \citenamefont {Zhang}(2011)}]{Qi11}%
  \BibitemOpen
  \bibfield  {author} {\bibinfo {author} {\bibfnamefont {X.-L.}\ \bibnamefont
  {Qi}}\ and\ \bibinfo {author} {\bibfnamefont {S.-C.}\ \bibnamefont {Zhang}},\
  }\href {\doibase 10.1103/RevModPhys.83.1057} {\bibfield  {journal} {\bibinfo
  {journal} {Rev. Mod. Phys.}\ }\textbf {\bibinfo {volume} {83}},\ \bibinfo
  {pages} {1057} (\bibinfo {year} {2011})}\BibitemShut {NoStop}%
\bibitem [{\citenamefont {Schnyder}\ \emph {et~al.}(2008)\citenamefont
  {Schnyder}, \citenamefont {Ryu}, \citenamefont {Furusaki},\ and\
  \citenamefont {Ludwig}}]{Schnyder08}%
  \BibitemOpen
  \bibfield  {author} {\bibinfo {author} {\bibfnamefont {A.~P.}\ \bibnamefont
  {Schnyder}}, \bibinfo {author} {\bibfnamefont {S.}~\bibnamefont {Ryu}},
  \bibinfo {author} {\bibfnamefont {A.}~\bibnamefont {Furusaki}}, \ and\
  \bibinfo {author} {\bibfnamefont {A.~W.~W.}\ \bibnamefont {Ludwig}},\ }\href
  {\doibase 10.1103/PhysRevB.78.195125} {\bibfield  {journal} {\bibinfo
  {journal} {Phys. Rev. B}\ }\textbf {\bibinfo {volume} {78}},\ \bibinfo
  {pages} {195125} (\bibinfo {year} {2008})}\BibitemShut {NoStop}%
\bibitem [{\citenamefont {Ryu}\ \emph {et~al.}(2010)\citenamefont {Ryu},
  \citenamefont {Schnyder}, \citenamefont {Furusaki},\ and\ \citenamefont
  {Ludwig}}]{Ryu10}%
  \BibitemOpen
  \bibfield  {author} {\bibinfo {author} {\bibfnamefont {S.}~\bibnamefont
  {Ryu}}, \bibinfo {author} {\bibfnamefont {A.~P.}\ \bibnamefont {Schnyder}},
  \bibinfo {author} {\bibfnamefont {A.}~\bibnamefont {Furusaki}}, \ and\
  \bibinfo {author} {\bibfnamefont {A.~W.~W.}\ \bibnamefont {Ludwig}},\ }\href
  {http://stacks.iop.org/1367-2630/12/i=6/a=065010} {\bibfield  {journal}
  {\bibinfo  {journal} {New J. Phys.}\ }\textbf {\bibinfo {volume} {12}},\
  \bibinfo {pages} {065010} (\bibinfo {year} {2010})}\BibitemShut {NoStop}%
\bibitem [{\citenamefont {Kitaev}(2009)}]{Kitaev09}%
  \BibitemOpen
  \bibfield  {author} {\bibinfo {author} {\bibfnamefont {A.}~\bibnamefont
  {Kitaev}},\ }\href {\doibase 10.1063/1.3149495} {\bibfield  {journal}
  {\bibinfo  {journal} {AIP Conf. Proc.}\ }\textbf {\bibinfo {volume} {1134}},\
  \bibinfo {pages} {22} (\bibinfo {year} {2009})}\BibitemShut {NoStop}%
\bibitem [{\citenamefont {Chiu}\ \emph {et~al.}(2016)\citenamefont {Chiu},
  \citenamefont {Teo}, \citenamefont {Schnyder},\ and\ \citenamefont
  {Ryu}}]{Chiu16}%
  \BibitemOpen
  \bibfield  {author} {\bibinfo {author} {\bibfnamefont {C.-K.}\ \bibnamefont
  {Chiu}}, \bibinfo {author} {\bibfnamefont {J.~C.~Y.}\ \bibnamefont {Teo}},
  \bibinfo {author} {\bibfnamefont {A.~P.}\ \bibnamefont {Schnyder}}, \ and\
  \bibinfo {author} {\bibfnamefont {S.}~\bibnamefont {Ryu}},\ }\href {\doibase
  10.1103/RevModPhys.88.035005} {\bibfield  {journal} {\bibinfo  {journal}
  {Rev. Mod. Phys.}\ }\textbf {\bibinfo {volume} {88}},\ \bibinfo {pages}
  {035005} (\bibinfo {year} {2016})}\BibitemShut {NoStop}%
\bibitem [{\citenamefont {Bianco}\ and\ \citenamefont
  {Resta}(2011)}]{Bianco11}%
  \BibitemOpen
  \bibfield  {author} {\bibinfo {author} {\bibfnamefont {R.}~\bibnamefont
  {Bianco}}\ and\ \bibinfo {author} {\bibfnamefont {R.}~\bibnamefont {Resta}},\
  }\href {\doibase 10.1103/PhysRevB.84.241106} {\bibfield  {journal} {\bibinfo
  {journal} {Phys. Rev. B}\ }\textbf {\bibinfo {volume} {84}},\ \bibinfo
  {pages} {241106} (\bibinfo {year} {2011})}\BibitemShut {NoStop}%
\bibitem [{\citenamefont {Prodan}\ \emph {et~al.}(2010)\citenamefont {Prodan},
  \citenamefont {Hughes},\ and\ \citenamefont {Bernevig}}]{Prodan10}%
  \BibitemOpen
  \bibfield  {author} {\bibinfo {author} {\bibfnamefont {E.}~\bibnamefont
  {Prodan}}, \bibinfo {author} {\bibfnamefont {T.~L.}\ \bibnamefont {Hughes}},
  \ and\ \bibinfo {author} {\bibfnamefont {B.~A.}\ \bibnamefont {Bernevig}},\
  }\href {\doibase 10.1103/PhysRevLett.105.115501} {\bibfield  {journal}
  {\bibinfo  {journal} {Phys. Rev. Lett.}\ }\textbf {\bibinfo {volume} {105}},\
  \bibinfo {pages} {115501} (\bibinfo {year} {2010})}\BibitemShut {NoStop}%
\bibitem [{\citenamefont {Prodan}(2010)}]{Prodan10_2}%
  \BibitemOpen
  \bibfield  {author} {\bibinfo {author} {\bibfnamefont {E.}~\bibnamefont
  {Prodan}},\ }\href {\doibase 10.1088/1367-2630/12/6/065003} {\bibfield
  {journal} {\bibinfo  {journal} {New J. Phys.}\ }\textbf {\bibinfo {volume}
  {12}},\ \bibinfo {pages} {065003} (\bibinfo {year} {2010})}\BibitemShut
  {NoStop}%
\bibitem [{\citenamefont {Prodan}(2011)}]{Prodan11}%
  \BibitemOpen
  \bibfield  {author} {\bibinfo {author} {\bibfnamefont {E.}~\bibnamefont
  {Prodan}},\ }\href {\doibase 10.1088/1751-8113/44/11/113001} {\bibfield
  {journal} {\bibinfo  {journal} {J. Phys. A: Math. Theor.}\ }\textbf {\bibinfo
  {volume} {44}},\ \bibinfo {pages} {113001} (\bibinfo {year}
  {2011})}\BibitemShut {NoStop}%
\bibitem [{\citenamefont {Loring}\ and\ \citenamefont
  {Hastings}(2010)}]{Loring10}%
  \BibitemOpen
  \bibfield  {author} {\bibinfo {author} {\bibfnamefont {T.~A.}\ \bibnamefont
  {Loring}}\ and\ \bibinfo {author} {\bibfnamefont {M.~B.}\ \bibnamefont
  {Hastings}},\ }\href {\doibase 10.1209/0295-5075/92/67004} {\bibfield
  {journal} {\bibinfo  {journal} {EPL}\ }\textbf {\bibinfo {volume} {92}},\
  \bibinfo {pages} {67004} (\bibinfo {year} {2010})}\BibitemShut {NoStop}%
\bibitem [{\citenamefont {Bianco}\ and\ \citenamefont
  {Resta}(2013)}]{Bianco13}%
  \BibitemOpen
  \bibfield  {author} {\bibinfo {author} {\bibfnamefont {R.}~\bibnamefont
  {Bianco}}\ and\ \bibinfo {author} {\bibfnamefont {R.}~\bibnamefont {Resta}},\
  }\href {\doibase 10.1103/PhysRevLett.110.087202} {\bibfield  {journal}
  {\bibinfo  {journal} {Phys. Rev. Lett.}\ }\textbf {\bibinfo {volume} {110}},\
  \bibinfo {pages} {087202} (\bibinfo {year} {2013})}\BibitemShut {NoStop}%
\bibitem [{\citenamefont {Mondragon-Shem}\ \emph {et~al.}(2014)\citenamefont
  {Mondragon-Shem}, \citenamefont {Hughes}, \citenamefont {Song},\ and\
  \citenamefont {Prodan}}]{MondragonShem14}%
  \BibitemOpen
  \bibfield  {author} {\bibinfo {author} {\bibfnamefont {I.}~\bibnamefont
  {Mondragon-Shem}}, \bibinfo {author} {\bibfnamefont {T.~L.}\ \bibnamefont
  {Hughes}}, \bibinfo {author} {\bibfnamefont {J.}~\bibnamefont {Song}}, \ and\
  \bibinfo {author} {\bibfnamefont {E.}~\bibnamefont {Prodan}},\ }\href
  {\doibase 10.1103/PhysRevLett.113.046802} {\bibfield  {journal} {\bibinfo
  {journal} {Phys. Rev. Lett.}\ }\textbf {\bibinfo {volume} {113}},\ \bibinfo
  {pages} {046802} (\bibinfo {year} {2014})}\BibitemShut {NoStop}%
\bibitem [{\citenamefont {Marrazzo}\ and\ \citenamefont
  {Resta}(2017)}]{Marrazzo17}%
  \BibitemOpen
  \bibfield  {author} {\bibinfo {author} {\bibfnamefont {A.}~\bibnamefont
  {Marrazzo}}\ and\ \bibinfo {author} {\bibfnamefont {R.}~\bibnamefont
  {Resta}},\ }\href {\doibase 10.1103/PhysRevB.95.121114} {\bibfield  {journal}
  {\bibinfo  {journal} {Phys. Rev. B}\ }\textbf {\bibinfo {volume} {95}},\
  \bibinfo {pages} {121114} (\bibinfo {year} {2017})}\BibitemShut {NoStop}%
\bibitem [{\citenamefont {Cardano}\ \emph {et~al.}(2017)\citenamefont
  {Cardano}, \citenamefont {D'Errico}, \citenamefont {Dauphin}, \citenamefont
  {Maffei}, \citenamefont {Piccirillo}, \citenamefont {de~Lisio}, \citenamefont
  {De~Filippis}, \citenamefont {Cataudella}, \citenamefont {Santamato},
  \citenamefont {Marrucci}, \citenamefont {Lewenstein},\ and\ \citenamefont
  {Massignan}}]{Cardano17}%
  \BibitemOpen
  \bibfield  {author} {\bibinfo {author} {\bibfnamefont {F.}~\bibnamefont
  {Cardano}}, \bibinfo {author} {\bibfnamefont {A.}~\bibnamefont {D'Errico}},
  \bibinfo {author} {\bibfnamefont {A.}~\bibnamefont {Dauphin}}, \bibinfo
  {author} {\bibfnamefont {M.}~\bibnamefont {Maffei}}, \bibinfo {author}
  {\bibfnamefont {B.}~\bibnamefont {Piccirillo}}, \bibinfo {author}
  {\bibfnamefont {C.}~\bibnamefont {de~Lisio}}, \bibinfo {author}
  {\bibfnamefont {G.}~\bibnamefont {De~Filippis}}, \bibinfo {author}
  {\bibfnamefont {V.}~\bibnamefont {Cataudella}}, \bibinfo {author}
  {\bibfnamefont {E.}~\bibnamefont {Santamato}}, \bibinfo {author}
  {\bibfnamefont {L.}~\bibnamefont {Marrucci}}, \bibinfo {author}
  {\bibfnamefont {M.}~\bibnamefont {Lewenstein}}, \ and\ \bibinfo {author}
  {\bibfnamefont {P.}~\bibnamefont {Massignan}},\ }\href {\doibase
  10.1038/ncomms15516} {\bibfield  {journal} {\bibinfo  {journal} {Nat.
  Commun.}\ }\textbf {\bibinfo {volume} {8}},\ \bibinfo {pages} {15516}
  (\bibinfo {year} {2017})}\BibitemShut {NoStop}%
\bibitem [{\citenamefont {Meier}\ \emph {et~al.}(2018)\citenamefont {Meier},
  \citenamefont {An}, \citenamefont {Dauphin}, \citenamefont {Maffei},
  \citenamefont {Massignan}, \citenamefont {Hughes},\ and\ \citenamefont
  {Gadway}}]{Meier18}%
  \BibitemOpen
  \bibfield  {author} {\bibinfo {author} {\bibfnamefont {E.~J.}\ \bibnamefont
  {Meier}}, \bibinfo {author} {\bibfnamefont {F.~A.}\ \bibnamefont {An}},
  \bibinfo {author} {\bibfnamefont {A.}~\bibnamefont {Dauphin}}, \bibinfo
  {author} {\bibfnamefont {M.}~\bibnamefont {Maffei}}, \bibinfo {author}
  {\bibfnamefont {P.}~\bibnamefont {Massignan}}, \bibinfo {author}
  {\bibfnamefont {T.~L.}\ \bibnamefont {Hughes}}, \ and\ \bibinfo {author}
  {\bibfnamefont {B.}~\bibnamefont {Gadway}},\ }\href {\doibase
  10.1126/science.aat3406} {\bibfield  {journal} {\bibinfo  {journal}
  {Science}\ }\textbf {\bibinfo {volume} {362}},\ \bibinfo {pages} {929}
  (\bibinfo {year} {2018})}\BibitemShut {NoStop}%
\bibitem [{\citenamefont {Huang}\ and\ \citenamefont
  {Liu}(2018{\natexlab{a}})}]{Huang18}%
  \BibitemOpen
  \bibfield  {author} {\bibinfo {author} {\bibfnamefont {H.}~\bibnamefont
  {Huang}}\ and\ \bibinfo {author} {\bibfnamefont {F.}~\bibnamefont {Liu}},\
  }\href {\doibase 10.1103/PhysRevB.98.125130} {\bibfield  {journal} {\bibinfo
  {journal} {Phys. Rev. B}\ }\textbf {\bibinfo {volume} {98}},\ \bibinfo
  {pages} {125130} (\bibinfo {year} {2018}{\natexlab{a}})}\BibitemShut
  {NoStop}%
\bibitem [{\citenamefont {Huang}\ and\ \citenamefont
  {Liu}(2018{\natexlab{b}})}]{Huang18_2}%
  \BibitemOpen
  \bibfield  {author} {\bibinfo {author} {\bibfnamefont {H.}~\bibnamefont
  {Huang}}\ and\ \bibinfo {author} {\bibfnamefont {F.}~\bibnamefont {Liu}},\
  }\href {\doibase 10.1103/PhysRevLett.121.126401} {\bibfield  {journal}
  {\bibinfo  {journal} {Phys. Rev. Lett.}\ }\textbf {\bibinfo {volume} {121}},\
  \bibinfo {pages} {126401} (\bibinfo {year} {2018}{\natexlab{b}})}\BibitemShut
  {NoStop}%
\bibitem [{\citenamefont {Focassio}\ \emph {et~al.}(2021)\citenamefont
  {Focassio}, \citenamefont {Schleder}, \citenamefont {Crasto~de Lima},
  \citenamefont {Lewenkopf},\ and\ \citenamefont {Fazzio}}]{Focassio21}%
  \BibitemOpen
  \bibfield  {author} {\bibinfo {author} {\bibfnamefont {B.}~\bibnamefont
  {Focassio}}, \bibinfo {author} {\bibfnamefont {G.~R.}\ \bibnamefont
  {Schleder}}, \bibinfo {author} {\bibfnamefont {F.}~\bibnamefont {Crasto~de
  Lima}}, \bibinfo {author} {\bibfnamefont {C.}~\bibnamefont {Lewenkopf}}, \
  and\ \bibinfo {author} {\bibfnamefont {A.}~\bibnamefont {Fazzio}},\ }\href
  {\doibase 10.1103/PhysRevB.104.214206} {\bibfield  {journal} {\bibinfo
  {journal} {Phys. Rev. B}\ }\textbf {\bibinfo {volume} {104}},\ \bibinfo
  {pages} {214206} (\bibinfo {year} {2021})}\BibitemShut {NoStop}%
\bibitem [{\citenamefont {Sykes}\ and\ \citenamefont
  {Barnett}(2021)}]{Sykes21}%
  \BibitemOpen
  \bibfield  {author} {\bibinfo {author} {\bibfnamefont {J.}~\bibnamefont
  {Sykes}}\ and\ \bibinfo {author} {\bibfnamefont {R.}~\bibnamefont
  {Barnett}},\ }\href {\doibase 10.1103/PhysRevB.103.155134} {\bibfield
  {journal} {\bibinfo  {journal} {Phys. Rev. B}\ }\textbf {\bibinfo {volume}
  {103}},\ \bibinfo {pages} {155134} (\bibinfo {year} {2021})}\BibitemShut
  {NoStop}%
\bibitem [{\citenamefont {Jezequel}\ \emph {et~al.}(2022)\citenamefont
  {Jezequel}, \citenamefont {Tauber},\ and\ \citenamefont
  {Delplace}}]{Jezequel22}%
  \BibitemOpen
  \bibfield  {author} {\bibinfo {author} {\bibfnamefont {L.}~\bibnamefont
  {Jezequel}}, \bibinfo {author} {\bibfnamefont {C.}~\bibnamefont {Tauber}}, \
  and\ \bibinfo {author} {\bibfnamefont {P.}~\bibnamefont {Delplace}},\
  }\href@noop {} {\bibfield  {journal} {\bibinfo  {journal} {arXiv:2203.17099}\
  } (\bibinfo {year} {2022})}\BibitemShut {NoStop}%
\bibitem [{\citenamefont {Wang}\ \emph {et~al.}(2022)\citenamefont {Wang},
  \citenamefont {Cheng}, \citenamefont {Liu}, \citenamefont {Liu},\ and\
  \citenamefont {Huang}}]{Wang22}%
  \BibitemOpen
  \bibfield  {author} {\bibinfo {author} {\bibfnamefont {C.}~\bibnamefont
  {Wang}}, \bibinfo {author} {\bibfnamefont {T.}~\bibnamefont {Cheng}},
  \bibinfo {author} {\bibfnamefont {Z.}~\bibnamefont {Liu}}, \bibinfo {author}
  {\bibfnamefont {F.}~\bibnamefont {Liu}}, \ and\ \bibinfo {author}
  {\bibfnamefont {H.}~\bibnamefont {Huang}},\ }\href {\doibase
  10.1103/PhysRevLett.128.056401} {\bibfield  {journal} {\bibinfo  {journal}
  {Phys. Rev. Lett.}\ }\textbf {\bibinfo {volume} {128}},\ \bibinfo {pages}
  {056401} (\bibinfo {year} {2022})}\BibitemShut {NoStop}%
\bibitem [{\citenamefont {Hannukainen}\ \emph {et~al.}(2022)\citenamefont
  {Hannukainen}, \citenamefont {Martinez}, \citenamefont {Bardarson},\ and\
  \citenamefont {Kvorning}}]{Hannukainen22}%
  \BibitemOpen
  \bibfield  {author} {\bibinfo {author} {\bibfnamefont {J.~D.}\ \bibnamefont
  {Hannukainen}}, \bibinfo {author} {\bibfnamefont {M.~F.}\ \bibnamefont
  {Martinez}}, \bibinfo {author} {\bibfnamefont {J.~H.}\ \bibnamefont
  {Bardarson}}, \ and\ \bibinfo {author} {\bibfnamefont {T.~K.}\ \bibnamefont
  {Kvorning}},\ }\href@noop {} {\bibfield  {journal} {\bibinfo  {journal}
  {arXiv:2207.01646}\ } (\bibinfo {year} {2022})}\BibitemShut {NoStop}%
\bibitem [{\citenamefont {Molignini}\ \emph {et~al.}(2022)\citenamefont
  {Molignini}, \citenamefont {Lapierre}, \citenamefont {Chitra},\ and\
  \citenamefont {Chen}}]{Molignini22_Chern_marker}%
  \BibitemOpen
  \bibfield  {author} {\bibinfo {author} {\bibfnamefont {P.}~\bibnamefont
  {Molignini}}, \bibinfo {author} {\bibfnamefont {B.}~\bibnamefont {Lapierre}},
  \bibinfo {author} {\bibfnamefont {R.}~\bibnamefont {Chitra}}, \ and\ \bibinfo
  {author} {\bibfnamefont {W.}~\bibnamefont {Chen}},\ }\href@noop {} {\bibfield
   {journal} {\bibinfo  {journal} {arXiv:2207.00016}\ } (\bibinfo {year}
  {2022})}\BibitemShut {NoStop}%
\bibitem [{\citenamefont {Chen}(2022)}]{Chen22_spin_Chern_marker}%
  \BibitemOpen
  \bibfield  {author} {\bibinfo {author} {\bibfnamefont {W.}~\bibnamefont
  {Chen}},\ }\href@noop {} {\bibfield  {journal} {\bibinfo  {journal}
  {arXiv:2207.02973}\ } (\bibinfo {year} {2022})}\BibitemShut {NoStop}%
\bibitem [{\citenamefont {von Gersdorff}\ \emph {et~al.}(2021)\citenamefont
  {von Gersdorff}, \citenamefont {Panahiyan},\ and\ \citenamefont
  {Chen}}]{vonGersdorff21_unification}%
  \BibitemOpen
  \bibfield  {author} {\bibinfo {author} {\bibfnamefont {G.}~\bibnamefont {von
  Gersdorff}}, \bibinfo {author} {\bibfnamefont {S.}~\bibnamefont {Panahiyan}},
  \ and\ \bibinfo {author} {\bibfnamefont {W.}~\bibnamefont {Chen}},\ }\href
  {\doibase 10.1103/PhysRevB.103.245146} {\bibfield  {journal} {\bibinfo
  {journal} {Phys. Rev. B}\ }\textbf {\bibinfo {volume} {103}},\ \bibinfo
  {pages} {245146} (\bibinfo {year} {2021})}\BibitemShut {NoStop}%
\bibitem [{\citenamefont {Chen}(2016)}]{Chen16}%
  \BibitemOpen
  \bibfield  {author} {\bibinfo {author} {\bibfnamefont {W.}~\bibnamefont
  {Chen}},\ }\href {http://stacks.iop.org/0953-8984/28/i=5/a=055601} {\bibfield
   {journal} {\bibinfo  {journal} {J. Phys. Condens. Matter}\ }\textbf
  {\bibinfo {volume} {28}},\ \bibinfo {pages} {055601} (\bibinfo {year}
  {2016})}\BibitemShut {NoStop}%
\bibitem [{\citenamefont {Chen}\ \emph {et~al.}(2016)\citenamefont {Chen},
  \citenamefont {Sigrist},\ and\ \citenamefont {Schnyder}}]{Chen16_2}%
  \BibitemOpen
  \bibfield  {author} {\bibinfo {author} {\bibfnamefont {W.}~\bibnamefont
  {Chen}}, \bibinfo {author} {\bibfnamefont {M.}~\bibnamefont {Sigrist}}, \
  and\ \bibinfo {author} {\bibfnamefont {A.~P.}\ \bibnamefont {Schnyder}},\
  }\href {http://stacks.iop.org/0953-8984/28/i=36/a=365501} {\bibfield
  {journal} {\bibinfo  {journal} {J. Phys. Condens. Matter}\ }\textbf {\bibinfo
  {volume} {28}},\ \bibinfo {pages} {365501} (\bibinfo {year}
  {2016})}\BibitemShut {NoStop}%
\bibitem [{\citenamefont {Chen}\ \emph {et~al.}(2017)\citenamefont {Chen},
  \citenamefont {Legner}, \citenamefont {R\"uegg},\ and\ \citenamefont
  {Sigrist}}]{Chen17}%
  \BibitemOpen
  \bibfield  {author} {\bibinfo {author} {\bibfnamefont {W.}~\bibnamefont
  {Chen}}, \bibinfo {author} {\bibfnamefont {M.}~\bibnamefont {Legner}},
  \bibinfo {author} {\bibfnamefont {A.}~\bibnamefont {R\"uegg}}, \ and\
  \bibinfo {author} {\bibfnamefont {M.}~\bibnamefont {Sigrist}},\ }\href
  {\doibase 10.1103/PhysRevB.95.075116} {\bibfield  {journal} {\bibinfo
  {journal} {Phys. Rev. B}\ }\textbf {\bibinfo {volume} {95}},\ \bibinfo
  {pages} {075116} (\bibinfo {year} {2017})}\BibitemShut {NoStop}%
\bibitem [{\citenamefont {Chen}\ and\ \citenamefont
  {Sigrist}(2019)}]{Chen19_AMS_review}%
  \BibitemOpen
  \bibfield  {author} {\bibinfo {author} {\bibfnamefont {W.}~\bibnamefont
  {Chen}}\ and\ \bibinfo {author} {\bibfnamefont {M.}~\bibnamefont {Sigrist}},\
  }\href@noop {} {\emph {\bibinfo {title} {Advanced Topological Insulators, Ch.
  7}}}\ (\bibinfo  {publisher} {Wiley-Scrivener},\ \bibinfo {year}
  {2019})\BibitemShut {NoStop}%
\bibitem [{\citenamefont {Chen}\ and\ \citenamefont
  {Schnyder}(2019)}]{Chen19_universality}%
  \BibitemOpen
  \bibfield  {author} {\bibinfo {author} {\bibfnamefont {W.}~\bibnamefont
  {Chen}}\ and\ \bibinfo {author} {\bibfnamefont {A.~P.}\ \bibnamefont
  {Schnyder}},\ }\href {\doibase 10.1088/1367-2630/ab2a2d} {\bibfield
  {journal} {\bibinfo  {journal} {New J. Phys.}\ }\textbf {\bibinfo {volume}
  {21}},\ \bibinfo {pages} {073003} (\bibinfo {year} {2019})}\BibitemShut
  {NoStop}%
\bibitem [{\citenamefont {Provost}\ and\ \citenamefont
  {Vallee}(1980)}]{Provost80}%
  \BibitemOpen
  \bibfield  {author} {\bibinfo {author} {\bibfnamefont {J.~P.}\ \bibnamefont
  {Provost}}\ and\ \bibinfo {author} {\bibfnamefont {G.}~\bibnamefont
  {Vallee}},\ }\href {https://projecteuclid.org:443/euclid.cmp/1103908308}
  {\bibfield  {journal} {\bibinfo  {journal} {Comm. Math. Phys.}\ }\textbf
  {\bibinfo {volume} {76}},\ \bibinfo {pages} {289} (\bibinfo {year}
  {1980})}\BibitemShut {NoStop}%
\bibitem [{\citenamefont {von Gersdorff}\ and\ \citenamefont
  {Chen}(2021)}]{vonGersdorff21_metric_curvature}%
  \BibitemOpen
  \bibfield  {author} {\bibinfo {author} {\bibfnamefont {G.}~\bibnamefont {von
  Gersdorff}}\ and\ \bibinfo {author} {\bibfnamefont {W.}~\bibnamefont
  {Chen}},\ }\href {\doibase 10.1103/PhysRevB.104.195133} {\bibfield  {journal}
  {\bibinfo  {journal} {Phys. Rev. B}\ }\textbf {\bibinfo {volume} {104}},\
  \bibinfo {pages} {195133} (\bibinfo {year} {2021})}\BibitemShut {NoStop}%
\bibitem [{\citenamefont {Panahiyan}\ \emph {et~al.}(2020)\citenamefont
  {Panahiyan}, \citenamefont {Chen},\ and\ \citenamefont
  {Fritzsche}}]{Panahiyan20_fidelity}%
  \BibitemOpen
  \bibfield  {author} {\bibinfo {author} {\bibfnamefont {S.}~\bibnamefont
  {Panahiyan}}, \bibinfo {author} {\bibfnamefont {W.}~\bibnamefont {Chen}}, \
  and\ \bibinfo {author} {\bibfnamefont {S.}~\bibnamefont {Fritzsche}},\ }\href
  {\doibase 10.1103/PhysRevB.102.134111} {\bibfield  {journal} {\bibinfo
  {journal} {Phys. Rev. B}\ }\textbf {\bibinfo {volume} {102}},\ \bibinfo
  {pages} {134111} (\bibinfo {year} {2020})}\BibitemShut {NoStop}%
\bibitem [{\citenamefont {Balian}\ and\ \citenamefont
  {Werthamer}(1963)}]{Balian63}%
  \BibitemOpen
  \bibfield  {author} {\bibinfo {author} {\bibfnamefont {R.}~\bibnamefont
  {Balian}}\ and\ \bibinfo {author} {\bibfnamefont {N.~R.}\ \bibnamefont
  {Werthamer}},\ }\href {\doibase 10.1103/PhysRev.131.1553} {\bibfield
  {journal} {\bibinfo  {journal} {Phys. Rev.}\ }\textbf {\bibinfo {volume}
  {131}},\ \bibinfo {pages} {1553} (\bibinfo {year} {1963})}\BibitemShut
  {NoStop}%
\bibitem [{\citenamefont {Volovik}(2009)}]{Volovik09}%
  \BibitemOpen
  \bibfield  {author} {\bibinfo {author} {\bibfnamefont {G.~E.}\ \bibnamefont
  {Volovik}},\ }\href@noop {} {\emph {\bibinfo {title} {The Universe in a
  Helium Droplet}}}\ (\bibinfo  {publisher} {Oxford University Press},\
  \bibinfo {year} {2009})\BibitemShut {NoStop}%
\bibitem [{\citenamefont {Bernevig}\ \emph {et~al.}(2006)\citenamefont
  {Bernevig}, \citenamefont {Hughes},\ and\ \citenamefont
  {Zhang}}]{Bernevig06}%
  \BibitemOpen
  \bibfield  {author} {\bibinfo {author} {\bibfnamefont {B.~A.}\ \bibnamefont
  {Bernevig}}, \bibinfo {author} {\bibfnamefont {T.~L.}\ \bibnamefont
  {Hughes}}, \ and\ \bibinfo {author} {\bibfnamefont {S.-C.}\ \bibnamefont
  {Zhang}},\ }\href {\doibase 10.1126/science.1133734} {\bibfield  {journal}
  {\bibinfo  {journal} {Science}\ }\textbf {\bibinfo {volume} {314}},\ \bibinfo
  {pages} {1757} (\bibinfo {year} {2006})}\BibitemShut {NoStop}%
\bibitem [{\citenamefont {K{\"o}nig}\ \emph {et~al.}(2007)\citenamefont
  {K{\"o}nig}, \citenamefont {Wiedmann}, \citenamefont {Br{\"u}ne},
  \citenamefont {Roth}, \citenamefont {Buhmann}, \citenamefont {Molenkamp},
  \citenamefont {Qi},\ and\ \citenamefont {Zhang}}]{Konig07}%
  \BibitemOpen
  \bibfield  {author} {\bibinfo {author} {\bibfnamefont {M.}~\bibnamefont
  {K{\"o}nig}}, \bibinfo {author} {\bibfnamefont {S.}~\bibnamefont {Wiedmann}},
  \bibinfo {author} {\bibfnamefont {C.}~\bibnamefont {Br{\"u}ne}}, \bibinfo
  {author} {\bibfnamefont {A.}~\bibnamefont {Roth}}, \bibinfo {author}
  {\bibfnamefont {H.}~\bibnamefont {Buhmann}}, \bibinfo {author} {\bibfnamefont
  {L.~W.}\ \bibnamefont {Molenkamp}}, \bibinfo {author} {\bibfnamefont {X.-L.}\
  \bibnamefont {Qi}}, \ and\ \bibinfo {author} {\bibfnamefont {S.-C.}\
  \bibnamefont {Zhang}},\ }\href {\doibase 10.1126/science.1148047} {\bibfield
  {journal} {\bibinfo  {journal} {Science}\ }\textbf {\bibinfo {volume}
  {318}},\ \bibinfo {pages} {766} (\bibinfo {year} {2007})}\BibitemShut
  {NoStop}%
\bibitem [{\citenamefont {Zhang}\ \emph {et~al.}(2009)\citenamefont {Zhang},
  \citenamefont {Liu}, \citenamefont {Qi}, \citenamefont {Dai}, \citenamefont
  {Fang},\ and\ \citenamefont {Zhang}}]{Zhang09}%
  \BibitemOpen
  \bibfield  {author} {\bibinfo {author} {\bibfnamefont {H.}~\bibnamefont
  {Zhang}}, \bibinfo {author} {\bibfnamefont {C.-X.}\ \bibnamefont {Liu}},
  \bibinfo {author} {\bibfnamefont {X.-L.}\ \bibnamefont {Qi}}, \bibinfo
  {author} {\bibfnamefont {X.}~\bibnamefont {Dai}}, \bibinfo {author}
  {\bibfnamefont {Z.}~\bibnamefont {Fang}}, \ and\ \bibinfo {author}
  {\bibfnamefont {S.-C.}\ \bibnamefont {Zhang}},\ }\href {\doibase
  10.1038/nphys1270} {\bibfield  {journal} {\bibinfo  {journal} {Nat. Phys.}\
  }\textbf {\bibinfo {volume} {5}},\ \bibinfo {pages} {438} (\bibinfo {year}
  {2009})}\BibitemShut {NoStop}%
\bibitem [{\citenamefont {Liu}\ \emph {et~al.}(2010)\citenamefont {Liu},
  \citenamefont {Qi}, \citenamefont {Zhang}, \citenamefont {Dai}, \citenamefont
  {Fang},\ and\ \citenamefont {Zhang}}]{Liu10}%
  \BibitemOpen
  \bibfield  {author} {\bibinfo {author} {\bibfnamefont {C.-X.}\ \bibnamefont
  {Liu}}, \bibinfo {author} {\bibfnamefont {X.-L.}\ \bibnamefont {Qi}},
  \bibinfo {author} {\bibfnamefont {H.}~\bibnamefont {Zhang}}, \bibinfo
  {author} {\bibfnamefont {X.}~\bibnamefont {Dai}}, \bibinfo {author}
  {\bibfnamefont {Z.}~\bibnamefont {Fang}}, \ and\ \bibinfo {author}
  {\bibfnamefont {S.-C.}\ \bibnamefont {Zhang}},\ }\href {\doibase
  10.1103/PhysRevB.82.045122} {\bibfield  {journal} {\bibinfo  {journal} {Phys.
  Rev. B}\ }\textbf {\bibinfo {volume} {82}},\ \bibinfo {pages} {045122}
  (\bibinfo {year} {2010})}\BibitemShut {NoStop}%
\bibitem [{\citenamefont {Chen}(2020)}]{Chen20_absence_edge_current}%
  \BibitemOpen
  \bibfield  {author} {\bibinfo {author} {\bibfnamefont {W.}~\bibnamefont
  {Chen}},\ }\href {\doibase 10.1103/PhysRevB.101.195120} {\bibfield  {journal}
  {\bibinfo  {journal} {Phys. Rev. B}\ }\textbf {\bibinfo {volume} {101}},\
  \bibinfo {pages} {195120} (\bibinfo {year} {2020})}\BibitemShut {NoStop}%
\bibitem [{\citenamefont {Su}\ \emph {et~al.}(1979)\citenamefont {Su},
  \citenamefont {Schrieffer},\ and\ \citenamefont {Heeger}}]{Su79}%
  \BibitemOpen
  \bibfield  {author} {\bibinfo {author} {\bibfnamefont {W.~P.}\ \bibnamefont
  {Su}}, \bibinfo {author} {\bibfnamefont {J.~R.}\ \bibnamefont {Schrieffer}},
  \ and\ \bibinfo {author} {\bibfnamefont {A.~J.}\ \bibnamefont {Heeger}},\
  }\href {\doibase 10.1103/PhysRevLett.42.1698} {\bibfield  {journal} {\bibinfo
   {journal} {Phys. Rev. Lett.}\ }\textbf {\bibinfo {volume} {42}},\ \bibinfo
  {pages} {1698} (\bibinfo {year} {1979})}\BibitemShut {NoStop}%
\bibitem [{\citenamefont {Kitaev}(2001)}]{Kitaev01}%
  \BibitemOpen
  \bibfield  {author} {\bibinfo {author} {\bibfnamefont {A.~Y.}\ \bibnamefont
  {Kitaev}},\ }\href {http://stacks.iop.org/1063-7869/44/i=10S/a=S29}
  {\bibfield  {journal} {\bibinfo  {journal} {Phys. Usp.}\ }\textbf {\bibinfo
  {volume} {44}},\ \bibinfo {pages} {131} (\bibinfo {year} {2001})}\BibitemShut
  {NoStop}%
\bibitem [{\citenamefont {Zhao}\ and\ \citenamefont {Wang}(2014)}]{Zhao14}%
  \BibitemOpen
  \bibfield  {author} {\bibinfo {author} {\bibfnamefont {Y.~X.}\ \bibnamefont
  {Zhao}}\ and\ \bibinfo {author} {\bibfnamefont {Z.~D.}\ \bibnamefont
  {Wang}},\ }\href {\doibase 10.1103/PhysRevB.90.115158} {\bibfield  {journal}
  {\bibinfo  {journal} {Phys. Rev. B}\ }\textbf {\bibinfo {volume} {90}},\
  \bibinfo {pages} {115158} (\bibinfo {year} {2014})}\BibitemShut {NoStop}%
\bibitem [{\citenamefont {Bernevig}\ and\ \citenamefont
  {Hughes}(2013)}]{Bernevig13}%
  \BibitemOpen
  \bibfield  {author} {\bibinfo {author} {\bibfnamefont {B.~A.}\ \bibnamefont
  {Bernevig}}\ and\ \bibinfo {author} {\bibfnamefont {T.~L.}\ \bibnamefont
  {Hughes}},\ }\href@noop {} {\emph {\bibinfo {title} {Topological Insulators
  and Topological Superconductors}}}\ (\bibinfo  {publisher} {Princeton
  University Press},\ \bibinfo {year} {2013})\BibitemShut {NoStop}%
\bibitem [{\citenamefont {Kane}\ and\ \citenamefont {Mele}(2005)}]{Kane05_2}%
  \BibitemOpen
  \bibfield  {author} {\bibinfo {author} {\bibfnamefont {C.~L.}\ \bibnamefont
  {Kane}}\ and\ \bibinfo {author} {\bibfnamefont {E.~J.}\ \bibnamefont
  {Mele}},\ }\href {\doibase 10.1103/PhysRevLett.95.226801} {\bibfield
  {journal} {\bibinfo  {journal} {Phys. Rev. Lett.}\ }\textbf {\bibinfo
  {volume} {95}},\ \bibinfo {pages} {226801} (\bibinfo {year}
  {2005})}\BibitemShut {NoStop}%
\end{thebibliography}%

\end{document}